\documentclass[fleqn,usenatbib]{mn2e}
\usepackage{tipa}
\usepackage{amsfonts}
\usepackage{bbm}
\usepackage{amsmath}
\usepackage{amssymb}
\usepackage{graphicx}

\def\gsim{\;\lower4pt\hbox{${\buildrel\displaystyle >\over\sim}$}\;}
\def\lsim{\;\lower4pt\hbox{${\buildrel\displaystyle <\over\sim}$}\;}
\def\grls{\;\lower4pt\hbox{${\buildrel\displaystyle >\over <}$}\;}

\title[Stability of a Dynamic Stellar Core Collapse]
{Perturbation Analysis of a General Polytropic\\
Homologously Collapsing Stellar Core}
\author[Y. Cao and Y.-Q. Lou]{Yi Cao$^1$\thanks{y-cao04@mails.tsinghua.edu.cn}
and Yu-Qing Lou$^{1,\ 2,\ 3}$\thanks{louyq@mail.tsinghua.edu.cn; \ \
    lou@oddjob.uchicago.edu }\\
    $^1$ Department of Physics and Tsinghua Centre for Astrophysics (THCA),
    Tsinghua University, Beijing 100084, China\\
    $^2$ Department of Astronomy and Astrophysics, the University of
    Chicago, 5640 S. Ellis Ave, Chicago, IL 60637, USA\\
    $^3$ National Astronomical Observatories, Chinese Academy of
    Sciences, A20, Datun Road, Beijing 100021, China
}
\date{Accepted 2009 August 21. Received 2009 July 27;
in original form 2009 April 25}

\begin{document}
\maketitle
\begin{abstract}
For dyanmic background models of Goldreich \& Weber and Lou \&
Cao, we examine three-dimensional perturbation properties of
oscillations and instabilities in a general polytropic
homologously collapsing stellar core of a relativistically hot
medium with a polytropic index $\gamma=4/3$. Perturbation
behaviours, especially internal gravity g$-$modes, depend on the
variation
of specific entropy in the collapsing core.
Among possible perturbations, we identify
acoustic p$-$modes and surface f$-$modes as well as internal
gravity g$^{+}-$modes and g$^{-}-$modes.
As in stellar oscillations of a static star, we define g$^{+}-$
and g$^{-}-$modes by the sign of the Brunt-V$\ddot{\rm
a}$is$\ddot{\rm a}$l$\ddot{\rm a}$ buoyancy frequency squared
${\cal N}^2$ for a collapsing stellar core. A new
criterion for the onset of convective instabilities is established
for a homologous stellar core collapse. We demonstrate that the
global energy criterion of Chandrasekhar is insufficient to
warrant the stability of general polytropic equilibria. We confirm
the acoustic p$-$mode stability of Goldreich \& Weber, even though
their p$-$mode eigenvalues appear in systematic errors. Unstable
modes include g$^{-}-$modes and sufficiently high-order
g$^{+}-$modes, both corresponding to convective core
instabilities.
Such instabilities occur before the stellar core bounce, in
contrast to instabilities in other models of supernova (SN)
explosions. The breakdown of spherical symmetry happens earlier
than expected in numerical simulations so far. The formation and
motion of the central compact object are speculated to be much
affected by such g$-$mode instabilities. By estimates of typical
parameters, unstable low-order $l=1$ g$-$modes may produce initial
kicks of the central compact object. Other high-order and
high-degree unstable g$-$modes may shred the nascent neutron core
into pieces without an eventual compact remnant (e.g. SN1987A).
Formation of binary pulsars and planets around neutron stars might
originate from unstable $l=2$ g$-$modes and high-order high-degree
g$-$modes, respectively.
\end{abstract}


\section{Introduction}

Supernovae (SNe), hypernovae and a few detected SNe associated
with long gamma-ray bursts (GRBs) serve as important cornerstones
of several major branches in astrophysics and cosmology. Physical
mechanisms and outcomes for such violent explosions of massive
stars have been actively pursued for decades.
Hydrodynamics and magnetohydrodynamics (MHD) together with
simplifying approximations and increasingly sophisticated
microphysics have been invoked to model various aspects of SNe in
both analytic treatments and numerical simulations. As nuclear
fuels eventually become insufficient in the stellar core, the
process of core-collapse SNe signaling the demise of massive
progenitors (e.g. red and blue giants) may be conceptually divided
into three stages of core collapse, rebound shock and neutrino
heating (e.g. Burrows et al. 1995; Janka \& M\"uller 1996).

The fortuitous detection of neutrinos from SN1987A (Hirata et
al. 1987; Bionta et al. 1987; Koshiba 2009 private
communications),
bolsters such a scenario framework in part or as a whole. Optical
observations before SN1987A revealed its progenitor as a blue
giant star in a mass range of $\sim 16-22M_\odot$ (e.g. Arnett et
al. 1989). At the time of SN1987A explosion, twenty neutrinos in
the energy range of $\sim 6-39$ MeV were intercepted within $\sim
12$ s, confirming the occurrence of neutronization. The timescale
of neutrino emissions was consistent with the prediction for
neutrino trapping inside an extremely dense collapsed core. The
total neutrino flux was consistent with energetic neutrinos
carrying off the binding energy during the core neutronization
(Chevalier 2009 private communications), even though no signals of
a neutron star (e.g. a pulsar) or a black hole are detected
(McCray 2009 private communications).

In spite of extensive research on analytic and numerical
studies of SNe over several decades (e.g. Goldreich \& Weber 1980
-- GW hearafter; Yahil 1983; Bruenn 1985; Bruenn 1989a, b; Herant
et al. 1995; Janka \& M${\rm\ddot{u}}$ller 1995, 1996; Fryer \&
Warren 2002, 2004; Blondin, Mezzacappa \& DeMarino 2003; Blondin
\& Mezzacappa 2006; Burrows et al. 2006, 2007a, b; Lou \& Wang
2006, 2007; Wang \& Lou 2007, 2008; Lou \& Cao 2008; Hu \& Lou
2009), a few major issues in the SN model development remain to be
explored (see Burrows et al. 2007 for a recent review). Among
these theoretical challenges, the dynamics of core-collapse stage
inside the progenitor and the possibility of convective
instabilities during this phase are the main thrust of this
paper.

The simplest hydrodynamic model to describe a core collapse
is a one-dimensional radial contraction with spherical symmetry
under the self-gravity. In analytical model analyses,
approximations to the equation of state (EoS) for gas medium are
necessarily introduced. It can be shown in statistical mechanics
(e.g. Huang 1987) that a relativistic hot Fermi gas with a
temperature much lower than the Fermi energy\footnote{The Fermi
energy is given by $E_F=[3h^3\rho Y_e/(4\pi
m_p)]^{1/3}c=30(\rho_{14}Y_e)^{1/3}$ MeV where $Y_e$ is the number
of electrons per baryon, $h$ is the Planck constant, $\rho$ is the
mass density, $m_p$ is the proton mass, $c$ is the speed of light,
$\rho_{14}$ is the mass density in unit of $10^{14}\hbox{ g
cm}^{-3}$.} can be modelled by a simple $\gamma=4/3$ polytropic
EoS with $\gamma$ being the polytropic index (i.e. the rest mass
of a single particle $\ll$ the kinetic energy of a particle $\ll$
the Fermi energy). This approximate EoS also gains support in
numerical simulations (e.g. Bethe et al. 1979; Hillebrandt, Nomoto
\& Wolff 1984; Shen et al. 1998). For instance, Bethe et al.
(1979) concluded that as the neutrino trapping occurs,
relativistic electrons, high-energy photons and neutrinos mainly
contribute to the total pressure within a collapsing stellar core
under gravity. As nuclei start to ``feel" each other at a high
density reaching up to $\sim 2.7\times 10^{14}{\rm g\ cm^{-3}}$,
the stiffness of nuclear matter adjusts the polytropic index
$\gamma$ to $\sim 2.5$.

There are two kinds of polytropic approximations. One is the
conventional polytropic EoS with $P=\kappa\rho^{\gamma}$ ($P$ and
$\rho$ are pressure and mass density, respectively) where
$\kappa$ remains constant in space and time. The other is a
general polytropic EoS where $\kappa$ remains constant along
streamlines, i.e.
\begin{eqnarray*}
\left(\frac{\partial}{\partial t}+\bf{
u}\cdot\nabla\right)\left(\frac{P}{\rho^\gamma}\right)=0\
,\label{funcofstat}
\end{eqnarray*}
where ${\bf u}$ is the bulk flow velocity. The former is actually
a special case of the latter, although we treat them separately.
It is easy to prove that for a relativistic hot Fermi gas with a
temperature much lower than the Fermi energy, the specific entropy
of the material is a function of $\log(P/\rho^\gamma)$ (see
footnote 1). In other words, the former EoS is interpreted as a
constant specific entropy while the latter bears a variable
distribution of specific entropy. The latter general polytropic
EoS obeys the conservation of specific entropy along streamlines.
The gas dynamics involves the competition between self-gravity and
pressure gradient force. Such models give rise to various
self-similar solutions by which the flow system partially loses
its memory of initial and boundary conditions (e.g. Larson 1969;
Penston 1969; Shu 1977; Cheng 1978; GW;
Yahil 1983; Suto \& Silk 1988; Lou \& Shen 2004; Yu et al. 2006;
Lou \& Wang 2006, 2007; Wang \& Lou 2007, 2008; Lou \& Cao 2008;
Hu \& Lou 2009). For core-collapse SNe, GW
studied the homologous stellar core collapse of a $\gamma=4/3$
conventional polytropic gas. They concluded that starting from a
static core, if the pressure is reduced by no more than $\sim
2.9\%$, the central core will evolve into a homologously
collapsing phase. This fraction is much less than $\sim 26\%$ as
indicated by Bethe et al. (1979). Yahil (1983) extended GW
analysis to $\gamma\le 4/3$ conventional polytropic cases and
noted the existence of an outer envelope moving inwards with a
supersonic speed. One important difference between GW and Yahil
(1983) is that solutions of the former have outer boundaries with
zero mass density there while those of the latter extend to
infinity (Lou \& Cao 2008).

With a similarity transformation (Fatuzzo et al.
2004), Lou \& Cao (2008) extended homologous core collapse to
general polytropic cases. This is a substantial theoretical
development of the model framework because several studies (e.g.
Bethe et al. 1979; Bruenn 1985, 1989b; Burrows et al. 2006;
Hillebrandt et al. 1984; Janka \& M$\ddot{\rm u}$ller 1995, 1996;
Shen et al. 1998) on EoS with microphysics during SNe do suggest
variable specific entropy depending on physical conditions,
including density, temperature and metallicity.

Meanwhile, extensive numerical simulations show that spherically
symmetric models cannot initiate SN explosions with an energy of
$\sim 10^{52}{\ \rm ergs}$ (e.g. Janka \& M$\ddot{\rm u}$ller
1995, 1996; Kitaura, Janka \& Hillebrandt 2006) partly because the
SN explosion energy appears insufficient and partly because
instabilities occur in multi-dimensional simulations. The role of
instabilities and symmetry breaking in SNe has now been emphasized
(e.g. Burrows 2000, 2006). Along this line, various instabilities
were proposed (e.g. Goldreich, Lai \& Sahrling 1996; Lai 2000; Lai
\& Goldreich 2000; Murphy, Burrows \& Heger 2004; Blondin et al.
2003; Blondin \& Mezzacappa 2006) and several mechanisms may
provide seed fluctuations before and during SN explosions (e.g.
Bazan \& Arnett 1998; Meakin \& Arnett 2006, 2007a, b). Prior to
the onset of a core collapse, the so-called
``$\epsilon$-mechanism" (e.g. Goldreich et al. 1996; Murphy et al.
2004) may lead to g$-$mode overstabilities in the progenitor, due
to the overreaction of nuclear processes against perturbations.

Lai \& Goldreich (2000) found an instability in the outer
supersonic envelope during the collapsing phase.
They performed both analytical and numerical irrotational
perturbation analysis using the conventional polytropic EoS for
collapsing solutions including EWCS of Shu (1977) and
post-collapse solution of Yahil (1983),
and found instabilities in supersonic regions. In their
derivation, the perturbed flow was assumed irrotational (i.e.
without vorticity) and thus g$-$modes should have been excluded.
In an example, they initiated their calculation with a g$-$mode
perturbation;
this initial condition appears to contradict the constraints of
their analytical derivations and numerical analysis.
After the emergence of a rebound shock, several instabilities have
been suggested. Intense convective motions may be sustained
outside the neutrino sphere (e.g. Herant, Benz \& Colgate 1992;
Herant et al. 1994). Standing accretion shock instability
(hereafter SASI; e.g. Foglizzo 2001) appears around $\sim 200$ ms
after the core bounce in some simulations (e.g. Blondin \&
Mezzacappa 2006; Blondin et al. 2003). Burrows et al. (2006,
2007a, b) proposed that $l=1$ g$-$modes at $\sim 500$ ms after the
stellar core bounce may serve as an agent to extract the
gravitational energy for the kinetic energy of SNe. The roles of
these instabilities are still hotly debated and a successful SN
explosion requires further explorations.

No significant core instabilities were reported for the
core-collapse phase. Linear stability analyses were performed by
GW, Lai (2000) and Lai \& Goldreich (2000) for certain dynamic
flows. GW perturbed their homologously collapsing solutions and
concluded that this collapse is stable for acoustic p$-$modes
with the g$-$modes being neutral convective modes. Lai (2000)
extended this acoustic stability analysis to solutions of Yahil
(1983) and found no unstable modes for $\gamma\simeq 4/3$ cases in
numerical explorations. Lai \& Goldreich (2000) studied the
stability of collapsing core and claimed that the core remains
stable in subsonic regions while the envelope becomes unstable in
supersonic regions.
All these stable core statements relies on the assumption of a
conventional polytropic EoS. Conventional polytropic gas flows and
perturbations correspond to a constant specific entropy and make
g$-$modes just neutral convective modes.

Our main theme is to examine stability properties of a
stellar core collapse with a variable specific entropy
distribution in a homologously collapsing model (Lou \& Cao 2008).
By numerical explorations, we classify various perturbation modes
including p$-$modes and g$-$modes (e.g. Cowling 1941),
some of which are oscillatory while others grow with time in power
laws. As the hydrostatic equilibrium is a limiting case of our
model, we find connection and evolution between perturbations in
progenitors of hydrostatic case and dynamic collapsing stage. The
most interesting result is that some stable g$-$modes in the
static case become unstable in the dynamically collapsing stage.
In particular, the stability of each mode now becomes sensitive to
the self-similar evolution of specific entropy. This instability
occurs during the core-collapse phase, neither before the core
collapse nor after the core bounce. We speculate implications of
such core instabilities during the collapse phase. For example,
the $l=1$ unstable g$-$mode may lead to the kick velocity of a
pulsar.

SN explosions involve a chain of physical processes. The
spherical symmetry may be destroyed by a series of instabilities
in the progenitor during the entire SN explosion. For example,
overstable g$-$modes of Goldreich et al. (1996) and Murphy et al.
(2004) may provide seed g$-$mode perturbations during the core
collapse. Perturbations of our model may connect to further
instabilities after the core bounce and the emergence of an
outgoing shock. Our model results are highly suggestive and can be
tested numerically.

This paper is structured as follows. Section 2 describes general
polytropic solutions of spherically symmetric homologous core
collapse, as a generalization of GW results. Homogeneous ordinary
differential equations (ODEs) for three-dimensional (3D) general
polytropic perturbations are obtained in Section 3. Numerical
results are analyzed in Section 4. We consider several aspects of
SNe in Section 5 and conclude in Section 6. Mathematical details
are summarized in Appendices A$-$C for the convenience of
reference.

\section{Spherical Homologous
Stellar Core Collapses}

Before starting the time-dependent 3D general polytropic
perturbation analysis, we first briefly summarize homologous core
collapse solutions with spherical symmetry of Lou \& Cao (2008),
making some notational adjustments in our model development for
the convenience of comparison with GW results. The nonlinear
partial differential equations (PDEs) for ideal hydrodynamics
are conservations of momentum and mass, Poisson equation for the
gravitational field and a general polytropic EoS, viz.
\begin{eqnarray}
\frac{\partial{\bf u}}{\partial t}+({\bf u}\cdot\nabla){\bf
u}=-\frac{1}{\rho}\nabla P
-\nabla\Phi\ ,\label{g1}\\
\frac{\partial\rho}{\partial
t}+\nabla\cdot\left(\rho{\bf u}\right)=0\ ,\label{g2}\\
\nabla^2\Phi=4\pi G\rho\ ,\label{g3}\\
P=\kappa({\bf r},\ t)\rho^\gamma\
,\qquad\quad\left(\frac{\partial}{\partial t}+{\bf
u}\cdot\nabla\right)\log\kappa=0\ ,\label{g4}
\end{eqnarray}
where ${\bf u}({\bf r},\ t)$, $P({\bf r},\ t)$, $\rho({\bf r},\
t)$ and $\Phi({\bf r},\ t)$ are bulk flow velocity, gas pressure,
mass density and gravitational potential of the flow system,
respectively and $G=6.67\times 10^{-8}$ cm$^3/$(g s$^2$) is the
universal gravitational constant, and $\gamma=4/3$ is the
polytropic index for a relativistically hot gas.

It is known that these ideal nonlinear hydrodynamic PDEs are
invariant under the time reversal operation,
\begin{eqnarray}
t\rightarrow-t,\quad {\bf u}\rightarrow-{\bf u},\quad
\rho\rightarrow \rho,\quad P\rightarrow
P,\quad\Phi\rightarrow\Phi\ .
\end{eqnarray}
This property enables us to use an outflow solution to also
describe a collapse process, which is very important to understand
homologous core collapse in terms of these expressions for the
time reversal invariance.

To generalize the analysis of GW model of spherical symmetry, we
introduce the following time-dependent spatial scale factor
$a(t)$,
\begin{eqnarray}
a(t)=\rho_c(t)^{-1/3}\left(\frac{\kappa_c}{\pi G}\right)^{1/2}\ ,
\end{eqnarray}
where time-dependent $\rho_c(t)$ and constant coefficient
$\kappa_c$ are the values of $\rho$ and $\kappa$ at the core
centre of a massive progenitor star (thus the subscript $_c$). The
dimensional vector radius $\vec{r}$ is scaled to a dimensionless
vector radius ${\bf x}=\vec{r}/a(t)$. Consistently, flow variables
of ${\bf u}(r,\ t)$, $P(r,\ t)$, $\rho(r,\ t)$ and $\Phi(r,\ t)$
are assumed to take on the following forms of
\begin{eqnarray}
{\bf u}=\dot{a}(t){\bf x}\ ,\label{eq8}\\
\rho=\rho_c(t)f^3(x)=\left(\frac{\kappa_c}{\pi G}
\right)^{3/2}a^{-3}f^3(x)\ ,\label{eq9}\\
P=\kappa\rho^{4/3}=\frac{\kappa_c^3}{(\pi G)^2}
a^{-4}g(x)f^4(x)\ ,\label{eq10}\\
\Phi=\frac{4}{3}\left(\frac{\kappa_c^3}{\pi
G}\right)^{1/2}a^{-1}\psi(x)\ ,\label{eq11}
\end{eqnarray}
where $\kappa(r,\ t)$ is prescribed in the form of $\kappa_c g(x)$
with a constant $\kappa_c$.
The central mass density $\rho_c(t)$ is proportional to
$a^{-3}(t)$ and varies with $t$. Substituting expressions
(\ref{eq8})$-$(\ref{eq11}) into nonlinear PDEs
$(\ref{g1})-(\ref{g4})$ under spherical symmetry with $x=r/a(t)$,
we reduce these nonlinear PDEs to a set of coupled nonlinear ODEs.
First, PDEs (\ref{g1}) and (\ref{g4}) are automatically satisfied,
the latter of which means that $g(x)$ can be of an arbitrary form.
For example, $g(x)=1$ brings our general polytropic EoS back to
the conventional polytropic EoS with a constant coefficient
$\kappa=\kappa_c$ studied by GW. Momentum equation (\ref{g1}) then
leads to
\begin{eqnarray}
-\left(\frac{\pi G}{\kappa_c^3}\right)^{1/2}a^2\ddot{a}=
\frac{1}{x}\left[\frac{1}{f^3}\frac{d}{dx}\left(gf^4\right)
+\frac{4}{3}\frac{d\psi}{dx}\right]\ ,
\end{eqnarray}
where the left-hand side (LHS) depends only on $t$ while the
right-hand side (RHS) depends only on $x$. For consistency, we
therefore need to set both sides equal to a constant $4\lambda/3$
and obtain two separate nonlinear ODEs, viz.
\begin{eqnarray}
-\left(\frac{\pi
G}{\kappa_c^3}\right)^{1/2}a^2\ddot{a}=\frac{4}{3}\lambda\ ,
\label{13}\\
\frac{d\psi}{dx}=\lambda
x-\frac{3}{4f^3}\frac{d}{dx}\left(gf^4\right)\ .\label{14}
\end{eqnarray}
ODE (\ref{13}) indicates that the spatial scale factor $a(t)$ is
either a constant independent of $t$ with $\lambda=0$ or a power
law of $t$ being proportional to $t^{2/3}$. Substituting equation
(\ref{14}) into Poisson equation (\ref{g3}), we derive an ODE for
$f(x)$, viz.
\begin{eqnarray}
\frac{1}{x^2}\frac{d}{dx}\left[\frac{x^2}{f^3}
\frac{d}{dx}\left(gf^4\right)\right]
+4f^3=4\lambda\ ,\label{equf}
\end{eqnarray}
which gives a profile of mass density $\rho$ by eq (\ref{eq9}) at
time $t$ and radius $r$ because $x$ is the independent similarity
variable combining $t$ and $r$ together. Once $f(x)$ is known,
other variables can all be readily derived. The special case of
$\lambda=0$ leads to the limit of general polytropic Lane-Emden
equation\footnote{The standard Lane-Emden equation is derived by
presuming a constant entropy in a gas sphere under the
self-gravity. For this entropy to be a function of $r$, we refer
to the resulting equilibrium equation as the general polytropic
Lane-Emden equation.
} (e.g. Eddington 1926; Chandrasekhar 1939) and $\lambda>0$
describes outflows or collapses by the time reversal operation.
For a necessary check, $g(x)=1$ in ODE (\ref{equf}) reduces to
equation (16) of GW precisely as expected.

The `boundary conditions' for second-order nonlinear ODE
(\ref{equf}) are as follows. The radial gradient of pressure should
vanish at the centre, i.e.
\begin{eqnarray}
\nabla P|_{x=0}=0\qquad\ \Rightarrow\qquad\ 4f'(0)+g'(0)=0\ ,
\end{eqnarray}
where the prime $'$ indicates the first derivative in terms of the
self-similar independent variable $x$. In order to obtain a
physically sensible solution of $f(x)$ related to the mass density
by algebraic expression (\ref{eq9}), we require an outer boundary
$x_b$ which is the smallest value of solutions $f(x)=0$. The
reason is, if no solution is found for $f(x)=0$ at a finite $x>0$,
it means that the system extends to infinity. Therefore, the
dimensional velocity ${\bf u}=2\vec{r}/(3t)$ will diverge towards
extremely large radii. Such a divergent flow velocity is
unacceptable in realistic astrophysical gas systems.

Hence, these boundary conditions determine a continuous range of
$\lambda$ values and
the maximum acceptable value of $\lambda$, denoted by $\lambda_M$
hereafter, corresponding to a solution $f(x)$ where $f'(x_b)=0$
also at the outer boundary $f(x_b)=0$. By numerical explorations,
solution $f(x)$ for $\lambda>\lambda_M$ does not go to zero at a
finite $x$ but oscillate with decreasing amplitude with increasing
$x$. Different profiles of $g(x)$ will lead to different values of
$\lambda_M$. For the special case of $g(x)=1$, the $\lambda$ range
is $0\le\lambda\le \lambda_M=0.00654376$ which was first
determined by GW and also confirmed by Lou \& Cao (2008) with a
corresponding $f_c\equiv f(0)=4.67047$.

The local polytropic sound speed is defined by
\begin{eqnarray}
V_s=\bigg(\frac{\partial P}{\partial
\rho}\bigg)^{1/2}\propto\big[g(x)\rho^{1/3}\big]^{1/2}\ .
\end{eqnarray}
In the theory of stellar oscillations, it is required that the
sound speed at the centre approaches a finite value (e.g. Unno et
al. 1979). We impose the same condition for a dynamic collapse.
Hence, $g'(0)=0$ and therefore $f'(0)=0$.

Once $f(x)$ is obtained, it is straightforward to calculate the
total enclosed mass of the collapsing core and the ratio between
the mean mass density and the central mass density in terms of
$f'(x_b)$ and $x_b$.
The total enclosed mass $M$ of the collapsing core is given by
\begin{eqnarray}
M\!\!\!\!\! &=&\!\!\!\!\!\int_0^{r_b}4\pi r^2\rho dr
=\frac{4}{3}\pi x_b^3\left(\frac{\kappa_c}{\pi
G}\right)^{3/2}\left[\lambda-\frac{3}{x_b}g(x_b)f'(x_b)\right]\nonumber\\
\!\!\!\!\!&=&\!\!\!\!\!\frac{4}{3}\pi
x_b^3\left(\frac{\kappa_c}{\pi
G}\right)^{3/2}\frac{\bar{\rho}}{\rho_c}\ ,
\end{eqnarray}
where the ratio between the mean mass density $\bar{\rho}$ and the
central mass density $\rho_c$ is readily identified with
\begin{eqnarray}
\frac{\bar{\rho}}{\rho_c}=\lambda-\frac{3}{x_b}g(x_b)f'(x_b)\ .
\end{eqnarray}
The mean mass density $\bar{\rho}$ varies with $t$ because $r_b$
decreases with $t$ for a homologous stellar core collapse. For
$\lambda=\lambda_M$, we have  $f'(x_b)=0$ and therefore
${\bar{\rho}}/{\rho_c}=\lambda_M$.

\section{Three-Dimensional Perturbations}

We now consider 3D general polytropic perturbations to the
background self-similar hydrodynamic collapse described in the
previous section in spherical polar coordinates $(r,\ \theta,\
\varphi)$. Flow variables including small perturbations are
assumed to bear the following forms of
\begin{eqnarray}
{\bf u}=\dot{a}(t){\bf x}+\frac{a}{t_{ff}}{\bf v}_1
(x,\ \theta,\ \varphi)\tau(t)\ ,\label{p1}\\
\rho=\left(\frac{\kappa_c}{\pi G}\right)^{3/2}a(t)^{-3}
f^3(x)[1+f_1(x,\ \theta,\ \varphi)\tau(t)]\ ,\label{p2}\\
P=\frac{\kappa_c^3}{(\pi G)^2}a(t)^{-4}g(x)f^4(x)[1
+\beta_1(x,\ \theta,\ \varphi)\tau(t)]\ ,\label{p3}\\
\Phi=\frac{4}{3}\left(\frac{\kappa_c^3}{\pi
G}\right)^{1/2}a(t)^{-1}[\psi(x)+\psi_1(x,\ \theta,\
\varphi)\tau(t)]\ ,\label{p4}
\end{eqnarray}
where the subscript $1$ indicates associations with perturbation
terms which are small compared to the background dynamic flow
variables and the free-fall timescale $t_{ff}$ is itself
time-dependent and is written in the specific form of
\begin{eqnarray}
t_{ff}&=&\left(\frac{4}{3}\pi G\rho_c\right)^{-1/2}\qquad\qquad
{\rm for\
all\ \lambda }\nonumber\\
&=&\left(\frac{9\lambda}{2}\right)^{1/2}t\qquad\qquad\qquad {\rm
for\ \lambda>0 }\ .\label{timedepend}
\end{eqnarray}
In expressions
(\ref{p1})$-$(\ref{p4}) above, $\tau(t)$ is a time-dependent
factor for perturbations and is assumed to bear the form of
\begin{eqnarray}
\tau(t)&=&\exp\left(p\int t_{ff}^{-1}dt\right)\qquad\qquad\quad
{\rm for\ all\
\lambda}\nonumber\\
&=&\exp\left[p\left(\frac{2}{9\lambda}\right)^{1/2}\ln t\right
]\qquad\quad {\rm for\ \lambda>0}\ ,\label{grow}
\end{eqnarray}
where the value of index parameter $p$ indicates either increase
or decrease as well as oscillations of perturbations relative to
the self-similar dynamic background flow.
Substituting expressions (\ref{p1})$-$(\ref{p4}) into nonlinear
PDEs (\ref{g1})$-$(\ref{g4}) with the standard linearization
procedure, we obtain equations governing 3D linear perturbations,
viz.
\begin{eqnarray}
m{\bf w}_1=-\frac{3}{4f^3}\left[\nabla\left(gf^4
\beta_1\right)-f_1\nabla\left(gf^4\right)\right]-\nabla\psi_1\
,\label{pf1}\\
ff_1+3{\bf w}_1\cdot\nabla f+f\nabla\cdot{\bf w}_1=0\ ,\label{pf2}\\
\nabla^2\psi_1=3f^3f_1\ ,\label{pf3}\\
g\left(\beta_1-\frac{4}{3}f_1\right)+{\bf w}_1\cdot\nabla g=0\
,\label{pf4}
\end{eqnarray}
for vector momentum equation, mass conservation, Poisson equation,
and specific entropy conservation along streamlines, respectively,
where the velocity perturbation ${\bf v}_1=p{\bf w}_1$ and
$m=p[p+(\lambda/2)^{1/2}]$. Therefore we have two values
$p=-(\lambda/8)^{1/2}\pm (\lambda/8+m)^{1/2}$ for a given
$\lambda$ and $m$. Our definition of $m$ here has an opposite sign
difference as compared to GW definition immediately after their
equation (27). For $\lambda/8+m<0$, we have a complex conjugate
pair for $p$ corresponding to oscillations, while for
$\lambda/8+m>0$, we have two real values of $p$ corresponding to
different perturbation growth rates for upper plus and lower minus
signs in $p$.

The angular components of $\psi_1$, $f_1$ and $\beta_1$ can be
readily separated out from the above perturbation equations by
spherical harmonics\footnote{To avoid notational confusions, we
use index $\mathfrak{m}$ for spherical harmonics
$Y_{l\mathfrak{m}}(\theta,\ \varphi)$ to distinguish from
eigenvalue parameter $m=p[p+(\lambda/2)^{1/2}]$.
The spherical harmonics is defined by
\begin{eqnarray*}
Y_{l\mathfrak{m}}(\theta,\varphi)=\bigg[\frac{(2l+1)}{4\pi}
\frac{(l-\mathfrak{m})!}{(l
+\mathfrak{m})!}\bigg]^{1/2}P_l^\mathfrak{m}
(\cos\theta)\exp(i\mathfrak{m}\varphi)\ ,\\
\mathfrak{m}=-l,\ -(l-1),\ \ldots,\ l-1,\ l\ ,
\end{eqnarray*}
where the associate Legendre polynomial
$P_l^\mathfrak{m}(\mathfrak{x})$ is defined as
\begin{eqnarray*}
P_l^\mathfrak{m}(\mathfrak{x})
=(1-\mathfrak{x}^2)^{|\mathfrak{m}|/2} \frac{d^{|\mathfrak{m}|}P_l
(\mathfrak{x})}{d\mathfrak{x}^{|\mathfrak{m}|}}\
\end{eqnarray*}
(e.g. Gupta 1978).}
and ${\bf w}_1$ takes the specific form of
\begin{eqnarray}
{\bf w}_1=\left({\bf e}_rw_r+{\bf e}_\theta
w_t\frac{\partial}{\partial\theta}+{\bf e}_\varphi
\frac{w_t}{\sin\theta}\frac{\partial}
{\partial\varphi}\right)Y_{l\mathfrak{m}}(\theta,\varphi)\ .
\end{eqnarray}
By this form of velocity perturbation, the radial component of
vorticity perturbation is zero, while the ${\bf e}_\theta$ and
${\bf e}_\phi$ components of vorticity perturbation do not vanish
in general. This allows the possible presence of g$-$mode
perturbations as well as convective motions and is distinctly
different from irrotational velocity perturbations of GW and Lai
\& Goldreich (2000). In the following analysis, $f_1$, $\beta_1$
and $\psi_1$ only describe the radial variations of respective
perturbation variables with the understanding that the relevant
angular parts involving the spherical harmonics
$Y_{l\mathfrak{m}}(\theta,\varphi)$ have been separated out. Using
equation (\ref{pf4}) and the angular (i.e. transverse) components
of equation (\ref{pf1}) to eliminate $\beta_1$ and $f_1$ in the
other three equations, one finally arrives at the following
fourth-order system of homogeneous linear ODEs for 3D general
polytropic perturbations, viz.
\begin{eqnarray}
\frac{1}{x^2}\frac{d}{dx}\left(x^2\frac{d\psi_1}
{dx}\right)-\frac{l(l+1)\psi_1}{x^2}
\qquad\qquad\qquad\qquad\qquad\nonumber\\
+\frac{3f^2}{g}\left(mxw_t-\frac{3}{4}f\frac{dg}{dx}w_r
+\psi_1\right)=0\ ,\label{pff1}\\
\left[m-\frac{9f}{16g}\left(\frac{dg}{dx}\right)^2
-\frac{9}{4}\frac{df}{dx}
\frac{dg}{dx}\right]w_r+\frac{3m}{4g}\frac{dg}{dx}xw_t\qquad\nonumber\\
-m \frac{d}{dx}\left(xw_t\right)+\frac{3}{4g}\frac{dg}{dx}\psi_1=0\
,\label{pff2}\\
\frac{f}{x^2}\frac{d}{dx}\left(x^2w_r\right)
-\frac{l(l+1)f}{x}w_t+3\frac{df}{dx}w_r\qquad\qquad\qquad\nonumber\\
-\frac{1}{g}
\left(mxw_t-\frac{3}{4}f\frac{dg}{dx}w_r+\psi_1\right)=0\
.\label{pff3}
\end{eqnarray}
Because of the background spherical symmetry,
3D general polytropic perturbations are degenerate with respect to
the azimuthal degree $\mathfrak{m}$ as expected (note that
$\mathfrak{m}$ and $m$ are two distinctly different parameters in
our notations).

Regular boundary conditions at both the centre and outer boundary
are required to keep perturbations physically sensible. They are
prescribed as follows.
\begin{eqnarray*}
\left\{
\begin{array}{c}
\psi_1\propto x^l\ ,\quad\quad w_r=lw_t\ \quad\quad{\rm
for\ } x\rightarrow 0^{+}\qquad\qquad\\
\\
\psi_1\propto x^{-(l+1)}\ ,\quad 3gw_rdf/dx-mxw_t=\psi_1\
\quad {\rm for\ }x=x_b\ .
\end{array}
\right.
\end{eqnarray*}
Boundary conditions at $x\rightarrow 0^{+}$ are imposed in order
to avoid singularity in perturbation solutions at the centre. The
other boundary condition at the moving radius of a collapsing core
is to require a zero Lagrangian pressure perturbation there, i.e.
${\partial P_1}/{\partial t}+({\bf u_1}\cdot\nabla)P_0=0$.

For the special case of $g(x)=1$ corresponding to a conventional
polytropic gas, these perturbation equations automatically reduce
to those of GW as expected.
Note that our Euler equation is written in a vector form because
the curl of the velocity perturbation field will not vanish for a
general polytropic gas and thus the stream function approach of GW
(whose gradient represents the velocity perturbation field) is not
sufficiently inclusive especially in view of possible g$-$mode
oscillations and convective instabilities. However, for a
conventional polytropic gas of constant $\kappa$ or $g(x)=1$, a
stream function can be defined and thus GW expressed Euler
equations in a scaler form by using such a stream function. They
perturb the stream function instead of the velocity field
directly.
More specifically, if the perturbation
stream function takes the form of $\Psi_1 Y_{l\mathfrak{m}}$, then
${\bf w}_1=\nabla\Psi_1$ gives
\begin{eqnarray}
xw_t=\Psi_1\ ,\qquad\qquad\qquad w_r=\frac{d\Psi_1}{dx}\
.\nonumber
\end{eqnarray}
This illustrates GW result being a special subcase of our more
general polytropic model description. In other words, GW consider
only perturbed potential flows without vorticity perturbations;
this approach suffices for purely acoustic oscillations. Likewise,
the irrotational perturbation flows of Lai \& Goldreich (2000)
should retain acoustic p$-$modes but exclude gravity g$-$modes and
convective motions. In our perturbation approach, vorticity
perturbations are present and all possible oscillations for
p$-$modes, f$-$modes, and g$-$modes are included in the model
consideration.

Several solution properties of this eigenvalue perturbation
problem can be demonstrated. For example, the eigenfunctions of
different eigenvalues $m$ are mutually orthogonal (see Appendix
A). In Appendix B, the eigenvalues and eigenfunctions can also be
written in terms of the variational principle (e.g. Chandrasekhar
1964). In particular, we demonstrate in Appendix C that the total
energy criterion of Chandrasekhar (1939) is not sufficient to
guarantee the stability of an equilibrium configuration in view of
the possible onset of convective instabilities for a variable
specific entropy distribution.

\section{Results of Perturbation Analysis}

In this section, 3D perturbation solutions are divided into
several classes analogous to the classification schemes of Cowling
(1941), Cox (1976) and Unno et al. (1979) for global stellar
oscillations of static spherical stars.

\subsection{A General Consideration }

Non-radial oscillation modes of a static spherical star have been
separated into different branches according to their respective
dominating restoring forces and their frequency ranges (e.g.
Cowling 1941).

The p$-$modes correspond to acoustic oscillation modes in which
pressure force is the major restoring force. Gravity modifies such
trapped sound waves in several ways. The characteristics of such
p$-$modes are: (1) The peaks of perturbation functions tend to
concentrate towards the outer envelope with increasing degree $l$
(i.e. larger $l$ values). (2) Their mode frequencies are
relatively high, compared with eigenfrequencies of other modes
such as f$-$modes and g$-$modes. (3) The more the number of radial
nodes in eigenfunctions, the higher the p$-$mode eigenfrequencies
and the more longitudinal the oscillations are. The radial
components of perturbed velocity dominate the oscillation in
high-degree (i.e. $l\gg 1$) modes.

Another type of oscillatory modes is the so-called internal
g$-$modes in which gravitational restoring force takes the
dominant role.
In contrast to p$-$modes, the g$-$mode characteristics are: (1)
The maxima of perturbation eigenfunctions bury deeply in the
stellar interior. (2) Their frequencies are relatively low. (3)
The frequency goes lower with the increase of the number of radial
nodes (the well-known anti-Sturmian property, e.g. Lou 1995).
As a limiting case, the frequency will approach zero and the
perturbation becomes nearly horizontal.

Between p$-$modes and g$-$modes, there exist the transitional
f$-$modes which have no nodes in both the mass density
perturbation and the radial component of velocity perturbation.
They are essentially surface modes in that perturbations have
evanescent behaviours beneath the surface layer. When the
perturbation degree $l$ becomes very large, the perturbation
concentrates around the surface layer. This mode is closely
related to the so-called Lamb waves (Lamb 1932; Lou 1990, 1991)
which propagate in the horizontal direction and vanish in the
vertical direction.
These f$-$modes may also be regarded as the lowest-order p$-$modes

More specifically, g$-$modes can be further divided into two
kinds, namely g$^{+}-$modes and g$^{-}-$modes (reading g$-$plus
modes and g$-$minus modes, respectively), according to whether the
eigenvalue $m$ is less or greater than zero. For global 3D
perturbations in static stars, the former class of modes is stable
while the latter class is unstable; the stability property of
these modes (especially the g$^{+}-$modes) is modified in
background dynamic collapses as discussed presently. The existence
of such two classes of g$-$modes depend on the square of the
so-called Brunt-V$\ddot{\rm a}$is$\ddot{\rm a}$l$\ddot{\rm a}$
buoyancy frequency ${\cal N}$, defined explicitly by
\begin{eqnarray}
{\cal N}^2\equiv {\cal G}\left(\frac{d\ln
\rho}{dr}-\frac{1}{\gamma}\frac{d\ln P}{dr}\right)\ \label{BVfreq}
\end{eqnarray}
where ${\cal G}$ is the magnitude of the local gravitational
acceleration and $\gamma$ is the polytropic index of
perturbations.

If ${\cal N}^2$ is positive everywhere, eigenvalues of $m$ are
always negative and only g$^{+}-$modes occur. If ${\cal N}^2$ is
negative everywhere, eigenvalues of $m$ are always positive and
only unstable g$^{-}-$modes occur. If ${\cal N}^2$ is positive in
a certain part of the system and negative in another part, both
types of g$-$modes may occur (e.g. Lebovitz 1965a, b, 1966 and Cox
1976 for stability properties of oscillations in static stars).
Note that the gravitational acceleration always points towards the
centre of the gas sphere (i.e. ${\cal G}>0$ by our notation).
Therefore, the sign of ${\cal N}^2$ is determined by the
expression in the parenthesis of definition (\ref{BVfreq}). In
fact, if the sign of the expression in the parenthesis is
negative, this region satisfies the Schwarzschild criterion for
convective instability (e.g. Lebovitz 1965a). Hence, the
g$^{-}-$modes represent convectively unstable modes. In the
simplest interpretation, ${\cal N}$ is the buoyancy frequency
associated with a perturbed parcel of fluid in a convectively
stable medium. Moreover, Scuflaire (1974) concluded for
oscillations in static stars that eigenfunctions of g$^{-}-$modes
can be oscillatory only in convectively unstable region (i.e.
${\cal N}^2<0$). On the other hand, eigenfunctions of
g$^{+}-$modes can be oscillatory only in radiative region (i.e.
${\cal N}^2>0$). We found similar features for perturbations in a
homologously collapsing stellar core by our extensive numerical
explorations.


Cowling (1941) showed and we readily confirm by definition
(\ref{BVfreq}) that for a conventional polytropic EoS
$P=\kappa\rho^\gamma$ with $\kappa$ being constant for both
dynamic core collapse and perturbations, the g$-$modes are simply
neutrally stable convective modes with ${\cal N}^2=0$ as noted by
GW. For this reason, GW investigated stability properties of
acoustic p$-$modes and suppressed vorticity perturbations in their
model analysis by introducing a stream function for velocity
perturbation. We note in passing that the situation with two
different polytropic indices for core collapse and perturbations
deserves a further investigation.

In contrast, our general polytropic EoS allows fairly free options
of specific entropy evolution along streamlines so that different
$g(x)$ gives different profiles of ${\cal N}^2$. Consequently,
g$-$modes can occur in a homologously collapsing stellar core and
can modify the convective instability criterion of such a dynamic
core collapse in a non-trivial manner.

\subsection{Numerical Explorations}

The numerical schemes we use to solve this perturbation problem
are described below. First, given a proper value of $\lambda$
parameter and a prescribed $g(x)$, we use an explicit fourth-order
Runge-Kutta scheme to numerically solve nonlinear ODE (\ref{equf})
for $f(x)$ and then determine all relevant background self-similar
dynamic flow variable profiles, i.e. $\psi(x)$ by first-order ODE
(\ref{14}), and then ${\bf u}(r,\ t)$, $P(r,\ t)$, $\rho(r,\ t)$
and $\Phi(r,\ t)$ by equations (\ref{eq8})$-$(\ref{eq11})
correspondingly. Having done this, we discretize self-similar
perturbation ODEs (\ref{pff1})$-$(\ref{pff3}) by using a proper
mesh in order to cast this eigenvalue problem of ODEs into a
matrix eigenvalue problem. Because small eigenvalues are
practically important, inverse iteration method is employed to
compute eigenvalues of $m$ accurately and to determine the
corresponding eigenfunctions (e.g. Wilkinson 1965).
For the purpose of checking, we substitute the obtained
eigenvalues and eigenfunctions into linear ODEs
(\ref{pff1})$-$(\ref{pff3}) to compute the residues which are
sufficiently small. This verification confirms the validity of our
numerical method. For double check, the Runge-Kutta shooting
method is also applied (i.e. starting numerical integrations near
both ends towards the centre) to verify the eigenvalues and
eigensolutions for perturbations. More specifically, this
perturbation problem involves undesirable diverging solutions
towards both ends. We use the numerical results obtained from the
matrix inverse iteration method at two respective points
sufficiently close both ends and integrate towards each other to
meet at a mid-point. This avoids the numerical difficulty of
diverging solutions towards both ends at $x=0$ and $x=x_b$.

Given a specified profile of $g(x)$ for the specific entropy
distribution with $g^{\prime}(x)=0$ at $x=0$ and a perturbation
degree $l$ value, numerical computations for 3D general polytropic
perturbations lead to families of eigenvalue $m(\lambda)$ curves.
One important feature for these curves as revealed by our
numerical explorations is that none of these families of
eigenvalue curves intersects with the $m=0$ line, i.e. each
eigenvalue branch $m(\lambda)$ remains either above or below $m=0$
line. Physically, this indicates that when $\lambda$ increases
from zero to the maximum $\lambda_M$, no oscillations initially
belonging to g$^{-}-$mode regime jump into g$^{+}-$mode regime
across the demarcation $m=0$. This is a general empirical
conclusion on the basis of our very extensive numerical
explorations.


The physically relevant form of $g(x)$, i.e. the radial evolution
of specific entropy along streamlines, requires a comprehensive
understanding of nuclear processes inside the high-density stellar
core under consideration. In order to effectively illustrate
essential features of our model through numerical explorations,
several possible trial distributions of specific entropy which
have relatively simple analytic forms of $g(x)$ are prescribed.
It should be emphasized that our model analysis regarding
g$-$modes does carry more general and important implications for
the stellar core collapse.


In the following, the conventional polytropic results of GW model
are first carefully examined by our approach and then three other
types of $g(x)$ are prescribed to explore stability properties of
p$-$modes, f$-$modes and g$-$modes.

\subsubsection{Comparisons with GW Model Results}

\begin{figure}
\includegraphics[width=0.5\textwidth]{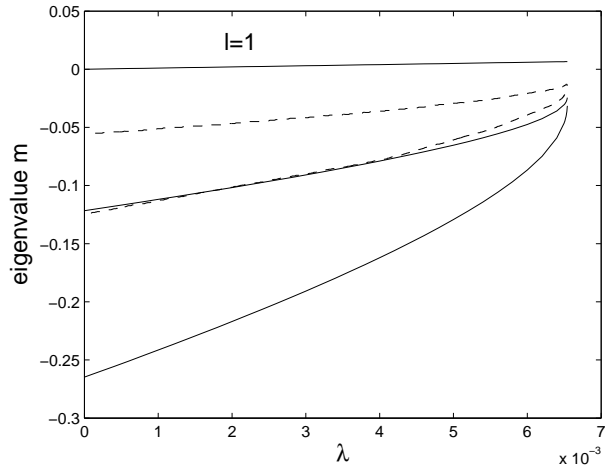}
\caption{The eigenvalues $m$ for the three lowest-order $l=1$
p$-$modes versus $\lambda$ variation in a conventional polytropic
stellar core (i.e. constant $\kappa$ or $g(x)=1$) in a homologous
collapse. The upper most one is an artificial mode branch
corresponding to a displacement of the coordinate origin. The
lower two branches are acoustic p$-$modes with zero and one node
in the radial velocity perturbation, respectively. The eigenvalues
of $m$ converge to $-25\lambda_M/8$ in the limit of
$\lambda\rightarrow\lambda_M$. The dashed curves represent GW
results for which we have already adjusted the sign difference
between eigenvalues of ours and those of GW. Our eigenvalues $m$
differ from those of GW significantly, even though the
corresponding eigenfunctions agree very well with those of GW.
}\label{fig1}
\end{figure}

\begin{figure}
\includegraphics[width=0.5\textwidth]{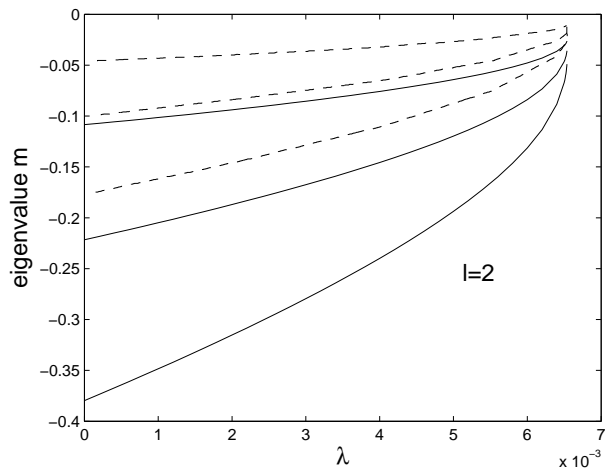}
\caption{The three branches (solid curves) of eigenvalues $m$ for
the three lowest-order $l=2$ p$-$modes versus $\lambda$ variation
in a conventional polytropic stellar core (i.e. constant $\kappa$
or $g(x)=1$) in a homologous collapse. They are all eigenvalues
for acoustic p$-$modes with zero, one and two nodes in the radial
velocity perturbation, respectively and converge to
$-25\lambda_M/8$ in the limit of $\lambda\rightarrow\lambda_M$.
The dashed curves represent GW eigenvalues for which we have
already adjusted the sign difference between eigenvalues of ours
and GW's. Again, our eigenvalues $m$ differ from those of GW
significantly, while the corresponding eigenfunctions agree very
well with those of GW.
}\label{fig2}
\end{figure}

\begin{figure}
\includegraphics[width=0.5\textwidth]{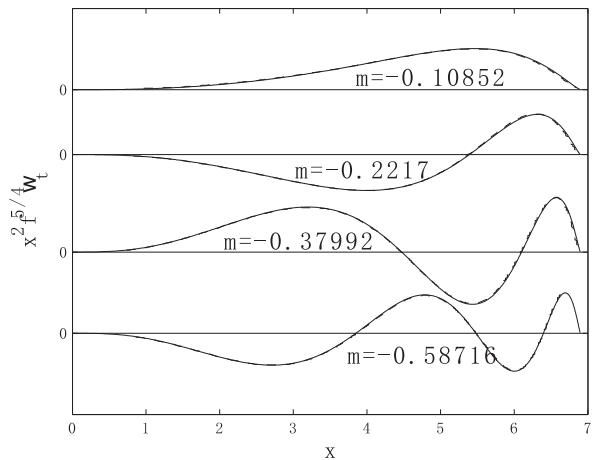}
\caption{
Acoustic p$-$mode eigenfunction profiles of
our $x^2f^{5/4}w_t$ (solid curves) which is equal to $rf^{5/4}w$
(dashed curves) of GW for $l=2$ and $\lambda=0$ (i.e. the static
conventional Lane-Emden sphere) are illustrated here for
comparison. The eigenfunctions of GW are displayed by dashed
curves; they are not readily distinguishable, because the
agreements of theirs and ours are extremely close.
We suspect that the eigenvalues of GW as shown in Figs. \ref{fig1}
and \ref{fig2} are in systematic errors.
}\label{fig3}
\end{figure}

\begin{figure}
\includegraphics[width=0.5\textwidth]{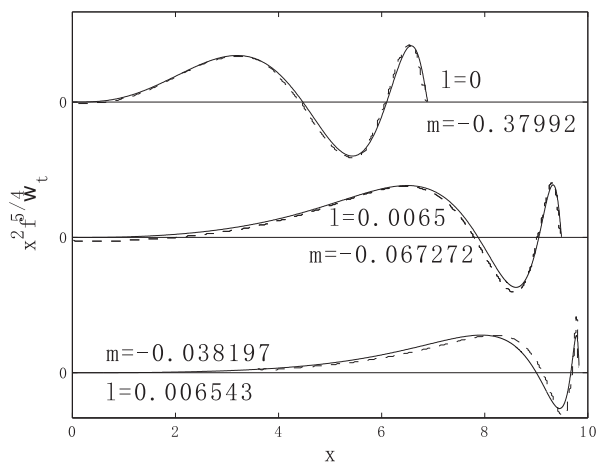}
\caption{
Here we show $l=2$ p$-$mode eigenfunctions with two radial nodes
in the transverse velocity perturbation for different values of
$\lambda$. These nodes concentrate towards the surface of the
dynamic stellar core as $\lambda$ approaches $\lambda_M$. The
dashed curves show GW results taken from their figure 3. Our
eigenfunctions agree well with those of GW but with a considerable
differences in the corresponding eigenvalues $m$ as displayed in
Fig. \ref{fig2}.
}\label{fig4}
\end{figure}

In order to compare with GW results for acoustic perturbations in
a conventional polytropic gas with $\gamma=4/3$, we simply set
$g(x)=1$ and let all terms containing derivatives of $g(x)$
vanish; this simplifies perturbation ODEs
(\ref{pff1})$-$(\ref{pff3}) considerably.
Comparing our results with those of GW, the eigenvalue curves
$m(\lambda)$ for the same oscillation p$-$modes do not coincide
with each other well. We have carefully checked our results in
several ways and suspect systematic computational errors in the
determination of eigenvalues\footnote{Our model calculations and
checks are carried out as follows. Using the matrix inverse
iteration procedure (e.g. Wilkinson 1965) which GW also used, we
obtain an eigenvalue with its corresponding eigenfunctions. We
then insert the eigenvalue and the eigenfunctions into nonlinear
ODEs (\ref{pff1})$-$(\ref{pff1}) to verify the results. Meanwhile,
given an initial value from the calculated eigenfunctions, we use
an explicit fourth-order Runge-Kutta scheme to solve ODEs
(\ref{pff1})$-$(\ref{pff1}) with the calculated eigenvalue again
to double-check the correctness of the results. In this
verification, regions near $x=0$ and the outer boundary $x=x_b$
are excluded to avoid diverging solutions there.
We then calculate and compare the counterpart results shown in the
figures of GW. After that, we find excellent coincidence in
eigenfunctions but systematic errors in eigenvalues between ours
and GW results.} by GW. Figs. \ref{fig1} and \ref{fig2} show the
first three branches for eigenvalues of $m$ versus $\lambda$ for
$l=1$ and $l=2$ p$-$modes (the branch with $m\gsim 0$ for $l=1$ is
an exception). For visual comparisons, GW eigenvalues are also
displayed by dashed curves in Figs \ref{fig1} and \ref{fig2}.
We have specifically compared the eigenfunctions of ours with
those of GW in Figs. \ref{fig3} and \ref{fig4}. Within numerical
errors, the corresponding eigenfunctions agree with each other
very well. We have carefully examined our computational procedures
and programs and make sure that the first three eigenvalues
correspond to the first three lowest-order p$-$mode oscillations
with $0$, $1$ and $2$ nodes along the radial direction. We suspect
systematic errors in the computational results of GW in terms of
their eigenvalues.

Figs. \ref{fig3} and \ref{fig4} show several eigenfunctions for
different eigenvalues of $m$ with specified $\lambda$ values for
the background dynamic collapse. In Figs. \ref{fig3} and
\ref{fig4}, dashed curves are those of GW results. The coincidence
of their curves and ours are very good. Fig. \ref{fig3} shows that
similar to classical non-radial oscillations of a static spherical
star, the $n$th lowest eigenvalue of a p$-$mode correspond to an
eigenfunction with $n-1$ number of nodes. Fig. \ref{fig4} shows
shape variations of eigenfunctions as $\lambda$ varies in
succession. The positive $m(\lambda)$ curve in the $l=1$ case is
an artificial one whose eigenfunction represents a movement of the
coordinate origin. Another exceptional curve noted by GW is for
the case of $l=0$, representing a different choice of the time
origin. Except for these two special cases, eigenvalue
$m(\lambda)$ curves of all families approach a common limit of
$-25\lambda_M/8$ in the limit of $\lambda\rightarrow\lambda_M$.
Curves of eigenvalue $m(\lambda)$ which are always smaller than
$-25\lambda_M/8$ must have such a limit. Otherwise, an essential
singularity of solutions appears for the limiting case of
$\lambda=\lambda_M$ (see GW for details).
Note that eigenfunctions of acoustic p$-$modes become concentrated
towards the surface as $\lambda$ approaches $\lambda_M$. This
singular behaviour of $\lambda_M$ should be regarded as an
artifact of the special mathematical character of this limiting
solution. In reality, the Lagrangian pressure perturbation
vanishes at the free surface. A finite sound speed at the surface
ensures that the normal modes are regular and the eigenvalues are
isolated.\footnote{As the limiting case of $\lambda_M$ only
belongs to a purely mathematical singularity and makes little
sense in physics, in the following discussion, we present common
features for general values of $\lambda$ without any special
attention towards this limiting case. }

\subsubsection{Further Numerical Explorations}

With the physical requirement of $g'(0)=0$, we consider three
types of variable $g(x)$ for numerical explorations. The first
type is an increasing function of $x$, the second type is a
decreasing function of $x$ and the last type has both increasing
and decreasing regions with increasing $x$. Referring to
definition (\ref{BVfreq}) of ${\cal N}^2$ for the buoyancy
frequency squared which serves as the criterion for the existence
of various g$-$modes, we find that the sign of ${\cal N}^2$
closely follows the sign of the first derivative of $g(x)$. Thus
in order to illustrate g$-$modes under various situations, such
three types of $g(x)$ are chosen as representatives. More
specifically, the three forms of $g(x)$ are prescribed as even
functions of $x$ and are given by
\begin{eqnarray}
g(x)=1+\frac{\varepsilon_1 x^2}{\varepsilon_1 x^2+1}\ ,\label{dg1}\\
g(x)=1-\frac{\varepsilon_2 x^2}{2(\varepsilon_2 x^2+1)}\ ,\label{dg2}\\
g(x)=1+\varepsilon_3 x^2\exp(-x^2/2)\ ,\label{dg3}
\end{eqnarray}
where $\varepsilon_{1}$, $\varepsilon_{2}$, and $\varepsilon_{3}$
are three parameters calibrating the values of respective
derivatives and thus the degree of variation for the dimensionless
$g(x)$.

\begin{figure}
\includegraphics[width=0.5\textwidth]{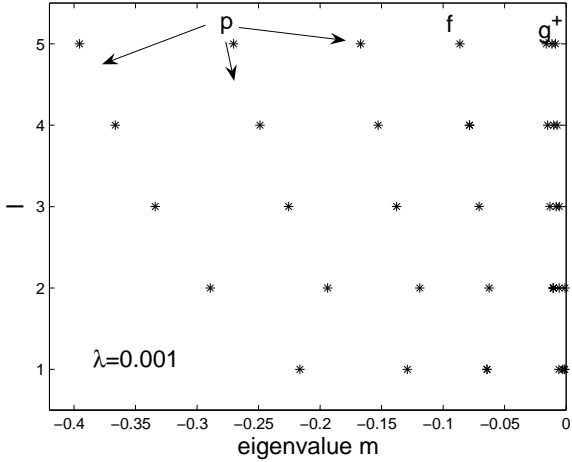}
\caption{Eigenvalues $m$ for various oscillation modes of several
lowest orders versus the angular spherical harmonic degree $l$ are
shown here. Three classes of modes, namely p$-$modes, f$-$modes
and g$^{+}-$modes, are identified according to the values of $m$
and the structures of their corresponding eigenfunctions.
Note that the specific entropy $g(x)$ is an increasing function of
$x$ in the form of expression (\ref{dg1}) with $\epsilon_1=0.01$
and parameter $\lambda=0.001$ is adopted. The g$^{-}-$modes do not
exist in this case. }\label{increase}
\end{figure}

\begin{figure}
\includegraphics[width=0.5\textwidth]{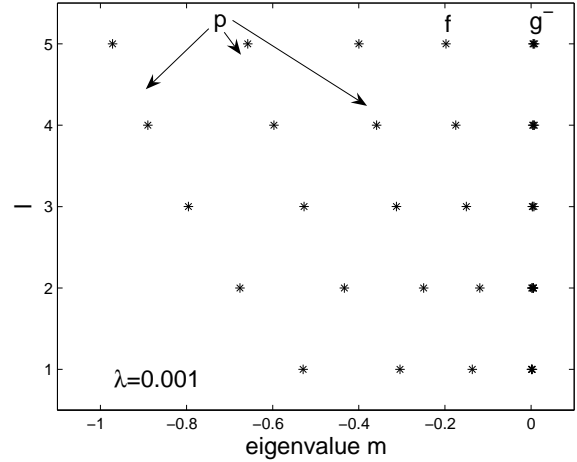}
\caption{Eigenvalues $m$ for various oscillation modes of several
lowest orders versus the angular spherical harmonic degree $l$ are
shown. Three classes of modes, namely p$-$modes, f$-$mode and
g$^{-}-$modes are identified according to the values of $m$ and
the structures of their corresponding eigenfunctions. Only
eigenvalues $m$ of g$^{-}-$modes are greater than zero. In this
case, the specific entropy $g(x)$ is a decreasing function of $x$
in the form of expression (\ref{dg2}) with $\epsilon_2=0.01$ and
parameter $\lambda=0.001$ is adopted. No g$^{+}-$modes appear in
this case. }\label{decrease}
\end{figure}

\begin{figure}
\includegraphics[width=0.5\textwidth]{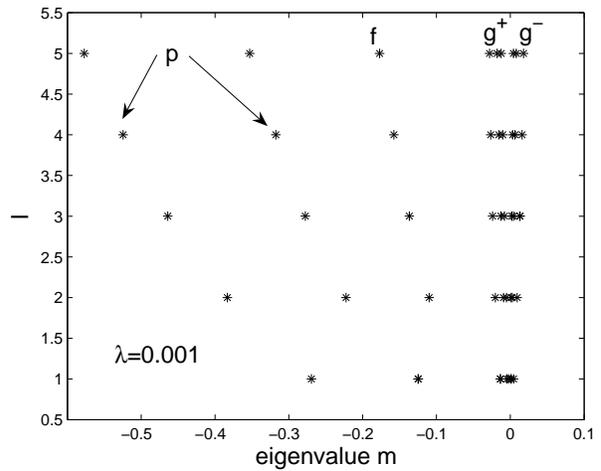}
\caption{Eigenvalues of $m$ parameter for various oscillation
modes of several lowest orders versus the angular spherical
harmonic degree $l$ are shown. Four classes of perturbation modes,
i.e. p$-$modes, f$-$modes, g$^{+}-$modes and g$^{-}-$modes, are
identified according to the values of $m$. Only eigenvalues of $m$
for convectively unstable g$^{-}-$modes are greater than zero.
Here, $g(x)$ takes the form of expression (\ref{dg3}) with
$\varepsilon_3=0.1$, neither purely increasing nor purely
decreasing, and parameter $\lambda=0.001$ is adopted.
}\label{varying}
\end{figure}

\begin{figure}
\includegraphics[width=0.5\textwidth]{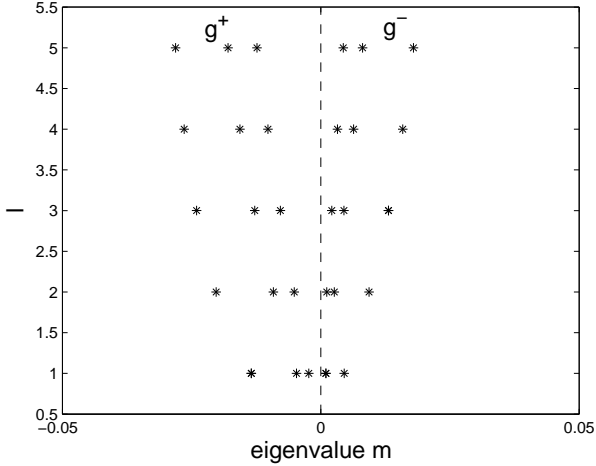}
\caption{The enlargement of region around the ordinate axis in
Fig. \ref{varying}. It is clear that g$^{+}-$modes and
g$^{-}-$modes are strictly divided by the dashed line $m=0$. The
g$^{+}-$ and g$^{-}-$modes are distinguished according to the sign
of $m$; the former corresponds to negative eigenvalues $m$ while
the latter corresponds to positive eigenvalues $m$. However, the
criterion for convective stability is $m<-\lambda/8$ in a
dynamically collapsing stellar core rather than $m<0$ for a static
background. It is therefore possible for g$^{+}-$modes to become
convectively unstable. }\label{amplification}
\end{figure}

\begin{figure}
\includegraphics[width=0.5\textwidth]{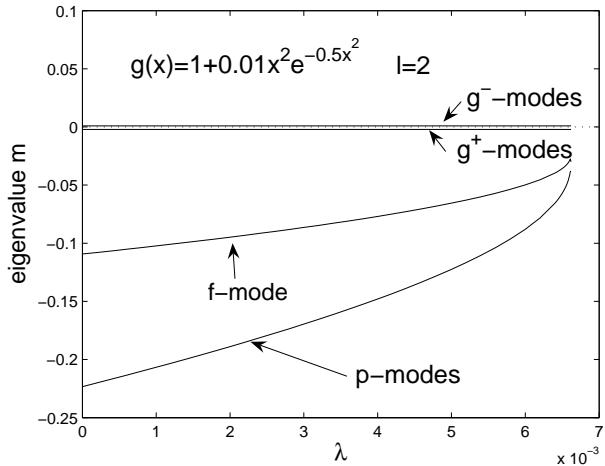}
\caption{Examples of eigenvalues $m$ of $l=2$ p$-$modes, f$-$modes,
g$^{+}-$mode and g$^{-}-$mode are shown in this case of
$g(x)=1+0.01x^2\exp(-x^2/2)$. }\label{lambda_of_m}
\end{figure}

\begin{figure}
\includegraphics[width=0.5\textwidth]{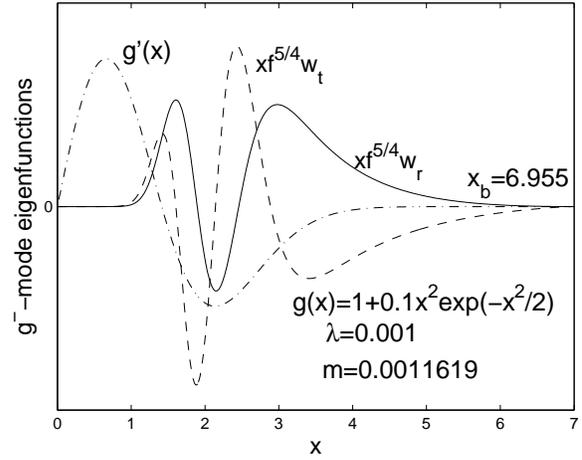}
\caption{As an example, eigenfunctions of a convectively unstable
g$^{-}-$mode for a $g(x)$ in the form of expression (\ref{dg3})
are shown. Relevant parameters are $\lambda=0.001$ and $l=2$ and
the form of $g(x)$ is shown in the figure.
The horizontal component $w_t$ (dashed curve) is larger than the
radial component $w_r$ (solid curve). Here, $g'(x)$ is shown by
the dash-dotted curve. The peaks of eigenfunctions concentrate in
the region of $g'(x)<0$ corresponding to ${\cal N}^2<0$.
}\label{example1}
\end{figure}

\begin{figure}
\includegraphics[width=0.5\textwidth]{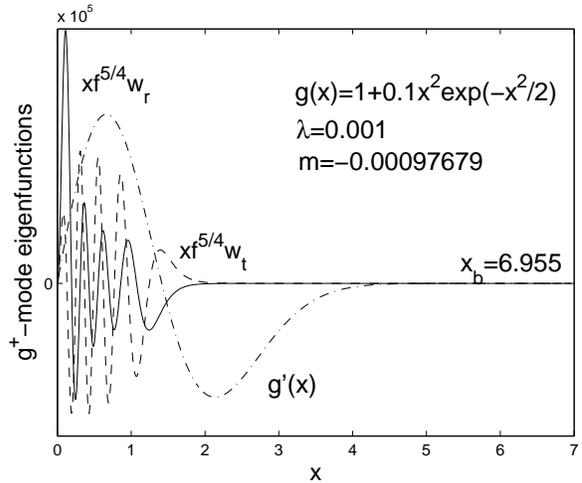}
\caption{As another example of illustration, eigenfunctions of a
stable g$^{+}-$modes for a $g(x)$ in the form of expression
(\ref{dg3}) are shown. Relevant parameters are $\lambda=0.001$ and
$l=2$ and the form of $g(x)$ is given in the figure.
The horizontal component $w_t$ (dashed curve) is larger than the
radial component $w_r$ (solid curve). The specific entropy $g'(x)$
is shown by dash-dotted curve. The peaks of eigenfunctions
concentrate in the central region, distinctive from g$^{-}-$modes
and p$-$modes. }\label{example2}
\end{figure}

\begin{figure}
\includegraphics[width=0.5\textwidth]{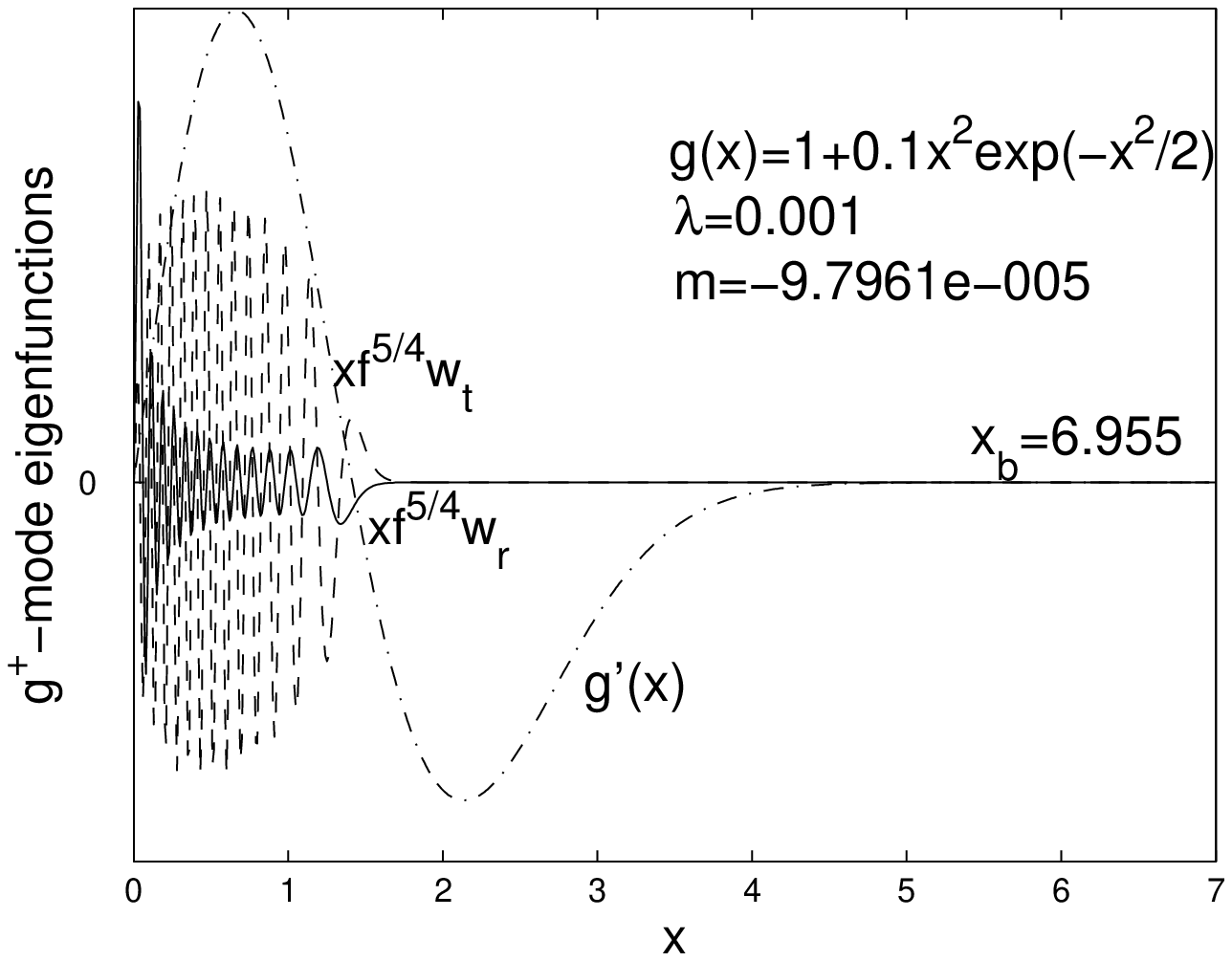}
\caption{As the third example of illustration, unstable
g$^{+}-$modes are shown with the eigenvalue $m$ exceeding
$-\lambda/8$. Relevant parameters and results are shown in the
figure. It is clear that $m+\lambda/8>0$ which corresponds to an
unstable mode. The dash-dotted curve is $g'(x)$. The concentration
of amplitude peaks within the region of $g'(x)>0$, i.e. ${\cal
N}^2>0$, and the negative value of eigenvalue $m$ both show that
it is a g$^{+}-$mode.}\label{example3}
\end{figure}

Several features of non-radial oscillations are contained in the
spectrum of eigenvalues $m$ versus spherical harmonic degree $l$
for a given value of $\lambda$ and a prescribed form of $g(x)$.
Figs. \ref{increase}, \ref{decrease} and \ref{varying}
schematically present the mode spectra for each type of specific
entropy $g(x)$ in equations (\ref{dg1}), (\ref{dg2}) and
(\ref{dg3}), respectively. Parameters $\varepsilon_1=0.01$,
$\varepsilon_2=0.01$ and $\varepsilon_3=0.1$ are adopted for our
numerical computations. Fig. \ref{amplification} zooms in the
neighborhood of the ordinate in Fig. \ref{varying}. Each asterisk
$*$ in the four figures represents computed eigenvalues $m$ (i.e.
the abscissa) for two or three lowest order modes versus the given
spherical harmonic degree $l$. As the background dynamic core
collapse is spherically symmetric, 3D perturbations are degenerate
with respect to the azimuthal degree $\mathfrak{m}$ (see footnote
2) as expected. Fig. \ref{lambda_of_m} further illustrates the
variation of $m$ versus $\lambda$ and reveals that no curve of
$m(\lambda)$ goes across the line of $m=0$.

By these figures,
it is clear to see the variation of eigenvalues with changing $l$.
The absolute values of $m$ for p$-$modes are largest. The higher
the order of a p$-$mode, the larger the absolute value of the
corresponding $m$. Moreover, we find that $m$ is always smaller
than $-\lambda/8$, which lead to an imaginary part in the power
index of time-dependent term $\tau(t)$ for perturbations.

Eigenvalues of g$-$modes, in contrast, lie in the regime of
smaller absolute value $m$, being further divided into two
sub-modes according to the signs of ${\cal N}^2$. Figs.
\ref{increase}, \ref{decrease} and \ref{varying} demonstrate the
criteria for the existence of g$^{-}-$modes and g$^{+}-$modes by
${\cal N}^2$, the square of the Brunt-V$\ddot{\rm a}$is$\ddot{\rm
a}$l$\ddot{\rm a}$ buoyancy frequency ${\cal N}$. In Fig.
\ref{increase} where the form of $g(x)$ keeps ${\cal N}^2$
positive, only g$^{+}-$modes exist and they are thus calculated;
g$^{-}-$modes do not exist as confirmed by our numerical
explorations. In Fig. \ref{decrease} where ${\cal N}^2$ is
negative, only unstable g$^{-}-$modes exist and they are thus
calculated; g$^{+}-$modes do not exist as confirmed by our
numerical explorations. In Fig. \ref{varying} where the sign of
${\cal N}^2$ varies in the region, both types of g$-$modes exist
and they are calculated by examples. Fig. \ref{amplification}
clearly shows the difference between g$^{-}-$modes and
g$^{+}-$modes in terms of eigenvalues of $m$.

%
Recalling the time-dependent factor $\tau(t)$ by definitions
(\ref{timedepend}) and (\ref{grow}) and the expression of
parameter $p=-(\lambda/8)^{1/2}\pm (m+\lambda/8)^{1/2}$, we find
that the criterion for convective stability changes with $\lambda$
value of a dynamic core collapse. As discussed presently, the
first term $-(\lambda/8)^{1/2}$ is interpreted as the gas
compression effect. Therefore for $m+\lambda/8>0$ of two real
values of $\pm (m+\lambda/8)^{1/2}$, one mode
decays with time $t$ while the other mode diverges with time $t$.
For $m+\lambda/8<0$, both modes oscillate with time $t$.
Consequently, the criterion for convective stability should be
$m<-\lambda/8$ rather than simply $m<0$. The most interesting
physical consequence is the appearance of unstable g$^{+}-$modes
of sufficiently high orders. Figs. \ref{example1}, \ref{example2}
and \ref{example3} demonstrate concrete examples of unstable
$g^{-}-$modes, stable $g^{+}-$modes and unstable $g^{+}-$modes,
respectively. The form of specific entropy $g(x)$ is described by
expression (\ref{dg3}) for $\epsilon_3=0.1$.

The unique f$-$modes (i.e. Lamb modes), shown in these figures and
treated as the lowest p$-$modes, exist only for $l\ge 2$. They
separate p$-$modes from g$-$modes in the sense that the
eigenvalues $m$ fall between the eigenvalues of p$-$modes and
those of g$-$modes. An important feature is that the
eigenfunctions of f$-$modes contain no nodes in the radial
component of velocity perturbation.

Non-radial oscillations of p$-$modes, f$-$modes and g$-$modes exist
for $l\ge 1$ (in particular f$-$modes requires $l\ge 2$).
Oscillations for $l=0$ modes are purely radial acoustic
oscillations as $Y_{00}=1/(2\pi)^{1/2}$. Since the horizontal
component vanishes, one can infer from the primary direction of
perturbation of different modes that only p$-$modes exist. Our
numerical results confirm this inference for a self-similar
dynamic core collapse in that only stable p$-$modes exist for
purely radial oscillations.

Moreover, several characteristic features of g$-$modes are
displayed in Figs. \ref{example1}, \ref{example2} and
\ref{example3}. These perturbations are primarily horizontal as
the horizontal component of perturbation velocity is usually
larger than the radial component of perturbation velocity. The
peaks of g$^{+}-$modes are trapped deeply in the stellar interior
of the collapsing stellar core while peaks of g$^{-}-$mode
eigenfunctions appear in the region of ${\cal N}^2<0$. Both have
much difference from acoustic p$-$modes in which the radial
component of perturbation dominates and oscillation peaks of
eigenfunctions appear towards the surface layer of a collapsing
stellar core.

\section{Model Considerations for SNe}

Homologous collapses have been invoked here to model the dynamic
phase of stellar core collapse prior to the emergence of a rebound
shock and the subsequent SN explosion. In numerical simulations
(e.g. Bruenn 1985) and theoretical studies (GW; Lai 2000), this
phase was claimed to be stable. We perform general polytropic 3D
perturbation analysis on such a homologous core collapse.
Analogous to stellar oscillations of a static star, 3D
perturbations during this dynamic core collapse can be classified
into several distinct modes. The most interesting and important
revelation is that convectively unstable g$-$modes appear under
generic conditions. When this happens, several physical
consequences for the remnant core follow, which may alter the
scenario for the breakup of spherical symmetry during the core
collapse.

\subsection{Radial Evolution of Specific Entropy}

In our model derivations and computations, we start from
assumptions of a general polytropic gas and the self-similar
collapse of such a gas sphere under self-gravity. For a homologous
core collapse, we find that the general polytropic EoS is
automatically satisfied. A radial evolution of specific entropy
$g(x)$ needs to be prescribed for a complete solution of a
homologous core collapse. Moreover, the instability of such a
dynamic core collapse depends on the radial evolution of specific
entropy. A brief review for the radial entropy distribution in the
core collapse phase deems necessary.


According to Bethe et al. (1979), entropy per nucleon (in unit of
Boltzmann's constant $k_B$) is $\sim 1-1.5$ at electron capture
where the mean electron number per baryon $Y_e$ is $\sim 0.31$.
This entropy value varies little such that a conventional
polytropic EoS was adopted by GW. Nevertheless, Bethe et al.
(1979) noted that multiple nuclear processes indeed change this
value from this range and this entropy variation serves as the
conceptual basis of our model. For example, when the breakup of
$^{56}$Fe occurs at a mass density of $\sim 5.9\times
10^{11}\hbox{ g cm}^{-3}$, entropy per nucleon varies from $\sim
0.5$ to $\sim 1.5$ at a temperature of $\sim 1-2$ MeV. While a
constant specific entropy may be an expedient approximation,
electron capture, neutrino trapping and other physical processes
certainly make the specific entropy vary with position and time.
It is this variation of specific entropy distribution that leads
to g$-$mode convective instabilities during a stellar core
collapse before the emergence of a rebound shock.

During the deleptonization process of a stellar core collapse, an
entropy evolution according to numerical simulations (e.g. Bruenn
1985; Janka et al. 1995, 1996; Burrows et al. 2006) might suggest
that specific entropy increases with increasing radius, as grossly
described by an increasing type of $g(x)$.
Bruenn (1985) displayed an increasing distribution of entropy
versus the enclosed mass. As the enclosed mass always increases
with radius, the specific entropy then ascends with increasing
radius. More evidence comes from simulations for the evolution
after the core bounce.
Several models illustrate the increasing trend of specific entropy
distribution with radius just after the core bounce (e.g. Janka et
al. 1995, 1996; Burrows et al. 2006). If the entropy distribution
varies not very much through the core bounce, then the specific
entropy distribution might also be an increasing function in the
core collapse stage, corresponding to an increasing trend of
$g(x)$ versus $x$.

In short, the analysis of Bethe et al. (1979) indicates that
entropy can vary in the core collapse phase while other numerical
simulations (e.g. Bruenn 1985, 1989a, b; Janka \& M$\ddot{\rm
u}$ller 1995, 1996; Burrows et al. 2006, 2007b) suggested an
increasing distribution of entropy during the phase of rebound
shock. Therefore, the overall distribution of entropy might be
assumed to increase with radius. Having said this, the possibility
cannot be ruled out that in parts of regions the entropy may
decrease with radius locally.

Bruenn (1985) pointed out that entropy decreases with density for
mass density $\lsim 10^{12}\ {\rm g\ cm}^{-3}$. A turning point
$\rho^{*}$ appears here, above which the entropy increases with
density. The reason is that at a certain nuclear density, the
energy level of $1f_{5/2}$ neutron shell in the nuclei becomes
filled and the electron capture by nuclei is no longer allowed
(e.g. Fuller et al. 1982). According to this result and the fact
that mass density is a decreasing function of radius, when
$\rho>\rho^{*}$ at a certain stage, the specific entropy
distribution may not be a monotonic increasing function of radius.
If this analysis reflects the physical reality, there are likely
some regions where the specific entropy decreases with radius.
Besides, detailed numerical simulations (e.g. Janka et al. 2007)
with more sophisticated EoS (e.g. Hillebrandt et al. 1984; Shen et
al. 1998) reveal that although the overall tendency is an outward
increase, the entropy distribution may have regions where the
entropy dips slightly.

The EoS and the specific entropy distribution remain open
questions. No definitive evidence requires necessarily a constant
or a monotonically increasing specific entropy distribution with
increasing radius during the core collapse phase. By our
computations, variable entropy regions certainly lead to unstable
g$-$modes with possibly rapid growth rates. These findings bear
important implications to perturbation growths during the core
collapse prior to the emergence of a rebound shock and SN
explosions.

\subsection{Definition of a Collapsing Inner Stellar Core}

For the type of SNe involving stellar core collapses,
a rebound shock emerges surrounding the centre because the inner
core is drastically compressed and stiffened obeying the EoS at
nuclear density. After such a core bounce, the central
neutron-rich core may become a proto-neutron star within a mass
range of $\sim 1-3M_\odot$ (e.g. Rhoades \& Ruffini 1974). One
expects that in the pre-collapse stage, there should exist a dense
central core with a comparable mass collapsing inwards to form
such a proto-neutron star. Some numerical simulations (e.g.
Woosley, Heger \& Weaver 2002) give the central core masses for
progenitors of various masses with different metallicities. The
core mass appears in the range of $\sim 1.2-1.8M_\odot$
($M_\odot=2\times 10^{33}$ g is the solar mass).

For a sudden core collapse, the outer layers of the progenitor may
not move in immediately (Burrows 2000). As the stellar core
rapidly contracts inwards, it may temporarily detach from outer
shells and evolve independently. In this sense, we conceive a
collapsing core under the self-gravity. Meanwhile, a quantitative
definition would be desirable. GW used the invariance of the inner
core mass to consider the maximum central pressure reduction
percentage from a $\lambda=0$ configuration to a homologous core
collapse. However, GW value of $\sim 3\%$ for the central pressure
reduction is much less than the result of $\sim 26\%$ in the
simulation of Bethe et al. (1979). GW defined a core mass as
\begin{eqnarray*}
M_{ic}=1.0449\left(\frac{\kappa_c}{\kappa_{c0}}\right)^{3/2}M_0\ ,
\end{eqnarray*}
where $M_0$ and $\kappa_{c0}$ are the enclosed mass and the value
of $\kappa_c$ for the $\lambda=0$ initial static Lane-Emden core
while the coefficient $1.0449$ comes from the variation of
$x_b^3\bar{\rho}/\rho_c$ as $\lambda$ varies from $0$ to the
maximum value $\lambda_M$. GW also noted that the inner core mass
in the computation of Van Riper (1978) is $\sim 30$\% larger than
that of their definition.

In reference to GW, we define the following inner core mass inside
the iron core of the progenitor. Given a form of specific entropy
evolution $g(x)$, we can determine $\lambda_M$ and then obtain by
what factor $c_1$, the value of $x_b^3\bar{\rho}/\rho_c$ increases
as $\lambda$ increases from $0$ to $\lambda_M$. As this inner
enclosed mass $M$ is proportional to
$x_b^3(\bar{\rho}/\rho_c)\kappa_c^{3/2}$, if we know the value of
$\kappa_c$ for the two cases of $\lambda=0$ and
$\lambda=\lambda_M$, the inner core mass is then defined by
\begin{eqnarray*}
M_{ic}=c_1\left(\frac{\kappa_c}{\kappa_{c0}}\right)^{3/2}M_0\ .
\end{eqnarray*}
This definition\footnote{In fact, there is yet another definition
for the inner core by Yahil (1983) in which the edge of the inner
core lies on the radius of maximum infall velocity. It is not
applicable here because Yahil's solution extends to infinity and
contains both the inner core and outer envelope while our solution
is valid up to a moving radius of zero mass density. }
  has an advantage that it allows arbitrary
  pressure reduction which triggers the SN explosion, though the inner
  core may be very small if the pressure reduction is great, especially
  in the GW cases of conventional polytropic EoS. In our general polytropic
  EoS characterized by a $g(x)$,
  the inner core mass can be larger according to this definition.
In our model development, we actually use parameter $\kappa_c$
which is the proportional coefficient between the pressure $P$ and
$\rho^{4/3}$ at the centre of a collapsing core to estimate the
inner core mass. According to the results shown in figure 4 of
Hillebrandt et al. (1984) which gave $P$ versus $\rho$ relations
for different entropies and values of the central entropy for
progenitors with different metallicities (e.g. Woosley et al.
2002), we infer parameter $\kappa_c$ to be $\gsim 10^{14}$ cgs
units. The total enclosed core mass is sensitive to the value of
$\kappa_c$ because of the power-law dependence of
$\kappa_c^{3/2}$. A form of specific entropy distribution
$g(x)=1+0.0001x^2\exp{(-x^2/2)}$
which deviates slightly from a constant is chosen here as a
demonstration. For $\kappa_c=3\times 10^{14}$ cgs units, the
enclosed core mass does not exceed $\sim 0.7M_\odot$; for
$\kappa_c=5\times10^{14}$ cgs units, the enclosed mass is around
$\sim 1.5M_\odot$; and for $\kappa_c=7\times 10^{14}$ cgs units,
the enclosed core mass can reach $\sim 2.5M_\odot$. As expected,
the central part gives the main contribution to the enclosed core
mass, i.e. materials are highly concentrated around the centre.
Typically, the chosen form of $g(x)$ does not change the enclosed
core mass very much unless it deviates from a constant
significantly.

\subsection{Properties of Various Perturbation Modes }
\subsubsection{Stable oscillations of p$-$modes and f$-$modes}

The high-frequency acoustic oscillations of p$-$modes and
f$-$modes may occur in any prescribed form of $g(x)$ and are
stable during the homologous core collapse. This conclusion here
is more general than that of GW for a conventional polytropic
stellar core collapse. In our model analysis and for convenience,
f$-$modes may be regarded as the lowest-order p$-$modes.
Hereafter, we do not distinguish the lowest order p$-$modes and
f$-$modes unless f$-$modes are necessarily emphasized. As
eigenvalues $m$ of p$-$modes decrease with increasing orders, if
the lowest order mode is stable, then all acoustic p$-$modes
remain stable. The local analysis in the
Wentzel-Kramers-Brillouin-Jeffreys (WKBJ) approximation suggests
the stability of high-order p$-$modes (GW).

A notable feature for perturbation modes in a dynamic core
collapse reveals that the time-dependent factor does not take on a
Fourier harmonic form $\exp(i\omega t)$ with $\omega$ being the
angular frequency of a perturbation. Instead, the temporal factor
acquires a power-law form, except for the limiting case of
$\lambda\rightarrow 0$ which consistently reduces to the
description of harmonic oscillations in a static general
polytropic sphere. For stable acoustic p$-$modes, all eigenvalues
$m$ found are negative and make $\lambda/8+m<0$ where $\lambda$
features the collapsing core. Therefore, the power-law index
parameter $\tau(t)$ has both real and imaginary parts. The real
part has a specific value of $-1/6$, corresponding to the
adiabatic amplification of acoustic waves due to compression of
gas collapsing towards the centre as noted by GW. The imaginary
part leads to a form of $\exp(i\zeta\ln t)$ with $\zeta=\pm
(-2m/\lambda-1/4)^{1/2}/3$ being a real number; this represents an
oscillation, but not in a conventional sinusoidal form of
harmonics.

For these oscillations of p$-$modes and f$-$modes during a dynamic
core collapse, analyses of GW and ours indicate stability. If one
includes effects of radiative losses and diffusive processes, then
some of these acoustic oscillations may become overstable, i.e.
oscillatory growths in a dynamic background. Such overstable
acoustic oscillations may serve as seed acoustic fluctuations for
the SASI (e.g. Foglizzo 2001) to operate during the subsequent
emergence of an outgoing rebound shock (e.g. Lou \& Wang 2006,
2007).

\subsubsection{Perturbations of g-modes and instabilities}

The most important results of our model analysis are that some
g$-$modes which were used to be regarded as convectively neutral
modes under the conventional polytropic approximation (GW)
can become unstable in a dynamic background with a general
polytropic core collapse under gravity. The occurrence of g$^{+}-$
and g$^{-}-$modes depends on the sign of ${\cal N}^2$ as
highlighted in Section 4.1.

In addition to the power-law factor $t^{-1/6}$ for oscillation
amplitudes by the core collapse compression, by using the modified
onset criterion of convective instabilities, i.e. $m>-\lambda/8$
(see Section 4.2.2), we find that g$^{-}-$modes and sufficiently
high-order g$^{+}-$modes are unstable during the core collapse.
Here, g$^{-}-$modes whose eigenvalues $m$ by definition are always
greater than zero are unstable. Table 1 provides information of
several illustrative examples of g$^{-}-$modes; eigenvalues $m$ of
g$^{+}-$modes approach $0$ when the order goes higher. As a
result, its value must exceed $-\lambda/8$ when the order is high
enough. Therefore high-order g$^{+}-$modes become unstable.
Consequently, unless for a constant $g(x)$,
unstable g$-$modes always exist, because at least one of the two
kinds of g$-$modes occurs for a variable distribution of specific
entropy.

It should be emphasized that numerical simulations so far have not
shown drastic growths of unstable convective modes found here.
Part of the reason is that there exists no systematic study on the
stability of the core collapse. One would have thought that
numerical truncation errors and errors in the determination of
thermodynamic variables from a tabulated EOS are considerable. It
would be highly desirable to further pursue this problem
numerically.

A physical scenario is advanced below for g$^{+}-$modes. When
g$^{-}-$modes do not occur, unstable g$^{+}-$modes may give rise
to convective instabilities during the core collapse. Initially,
the inner core of a progenitor remains in a $\lambda=0$
configuration for which g$^{+}-$modes oscillate stably. When an
insufficient nuclear energy supply triggers a reduction of core
pressure, the inner core begins to collapse homologously.
Sufficiently high order g$^{+}-$modes become unstable, starting to
grow and break the spherical symmetry of the collapsing core. Such
convective instability is limited during the core collapse because
the growth rate does not exceed $t^{-1/6}\propto a^{-1/4}$ because
g$^{+}-$modes are defined by $m<0$. If the spatial scale of the
core shrinks by a factor of $\sim 10^3$, perturbation grows by a
factor of $\sim 6$. This is comparable to the compression effect
of a collapsing core.

We speculate that such g$-$mode instabilities under favorable
conditions might break up a collapsing core of high density and
influence the formation of the central compact object and its
companions, such as binary pulsars (e.g. Hulse \& Taylor 1975) and
planets around neutron stars (e.g. Bailes, Lyne \& Shemar 1991;
Wolszczan \& Frail 1992).

The importance of asymmetry has been emphasized recently for the
core bounce in a progenitor star prior to SN explosions.
So far, one-dimensional model cannot lead to a successful SN
explosion. Another consensus is that the neutrino heating
mechanism alone also fails to produce a SN explosion as the energy
of $\gtrsim 10^{50}\ {\rm ergs}$ appears insufficient by one or
two orders of magnitude (e.g. Kirauta et al. 2006; Burrows et al.
2007a). In recent years, physical mechanisms involving two kinds
of fluctuations have been proposed to effectively extract the
available gravitational energy to power SN explosions. One is the
SASI process,
which invokes acoustic fluctuations
to transfer energy. The other process relying on $l=1$ g$-$modes
appears at several hundred microseconds after the core bounce as
simulated by Burrows et al. (2006, 2007a, b). These two energy
transfer processes destroy the spherical symmetry and make SN
explosions anisotropic.

Regarding the origin of such oscillatory modes or fluctuations,
our proposed instabilities which take place during the core
collapse phase before the emergence of a rebound shock should have
already destroyed the spherical symmetry. The condition for the
occurrence of such instabilities appears generic, only requiring a
variable radial specific entropy distribution.
Physically, the g$^{-}-$modes and unstable g$^{+}-$modes lead to
convections. By definition (\ref{BVfreq}) for the
Brunt-V$\ddot{\rm a}$is$\ddot{\rm a}$l$\ddot{\rm a}$ frequency
${\cal N}$, inequality ${\cal N}^2>0$ is equivalent to the
Schwarzschild criterion for convective stability in a star. The
global eigenfunction of a g$^{-}-$mode describes convective
motions in a homologous core collapse. For a given mode, we can
calculate the time evolution of the perturbation. Table
\ref{table1} gives an example of the power indices for the lowest
order $l=2$ mode under the prescribed form (\ref{dg3}) of
$g(x)=1-0.1x^2\exp(-x^2)$. According to Table \ref{TB1}, the
power-law index varies in a wide range, allowing for both fast and
slow perturbation growths, indicating the trend that instabilities
go into the nonlinear regime.

\begin{table}
\begin{center}
\caption{The power-law index  $-[1/36+2m/(9\lambda)]^{1/2}$ in the
time-dependent factor $\tau(t)$ (see eqns (\ref{timedepend}) and
(\ref{grow})) for the lowest order $l=2$ unstable g$^{-}-$modes
varies with $\lambda$ values for the dynamic core collapse. There
are two values of $p$ resulting from one eigenvalue $m$; as the
plus sign corresponds to stable modes, we only list the index of
unstable modes with minus signs in calculating index $p$. In
calculations of this example, we find that the decrease of
eigenvalue $m$ does not exceed $0.6\%$ when $\lambda$ increases
from $0$ (static core) to the maximum value $\lambda_M=0.00607$.
The specific entropy distribution $g(x)=1-0.1x^2\exp(-x^2)$ is
adopted according to expression (\ref{dg3}). }\label{table1}
\begin{tabular}{cc}
 $\lambda$ & $-[1/36+2m/(9\lambda)]^{1/2}$\\ \hline
 0.001     & $-1.663$\\
 0.002     & $-1.181$\\
 0.003     & $-0.969$\\
 0.004     & $-0.843$\\
 0.005     & $-0.757$\\
 0.006     & $-0.694$\\
 $\lambda_M=0.00607$ & $-0.690$\\ \hline
 \label{TB1}
\end{tabular}
\end{center}
\end{table}

\subsection{Comparisons with Stellar Oscillations}

Apparent similarities exist in parallel between stellar
oscillations of a static star and 3D general polytropic
perturbations in a homologous core collapse. The classification of
different oscillatory modes are introduced in reference to stellar
oscillations and the degeneracy with respect to azimuthal degree
$\mathfrak{m}$ is expected (see footnote 2). Acoustic p$-$modes
exist for all values of perturbation degree $l$ and remain stable.
The acoustic f$-$modes exist for $l\ge 2$ and also remain stable.
The criterion for the existence of two different types of
g$-$modes remains the same, by using the sign of the square of the
Brunt-V$\ddot{\rm a}$is$\ddot{\rm a}$l$\ddot{\rm a}$ buoyancy
frequency ${\cal N}^2$ defined by expression (\ref{BVfreq}).
Similar to stellar oscillations, amplitudes of g$-$modes
concentrate around the core while those of p$-$modes approach the
outer layer especially for high-degree $l$ acoustic oscillations.
More specifically, peak amplitudes of g$^{-}-$modes appear in the
radial regions with ${\cal N}^2<0$.

Meanwhile, notable differences from stellar oscillations also
arise in our perturbation analysis. As the most important
features, the onset criterion for convective instability is
modified in reference to the well-known Schwarzschild criterion.
Consequently, not only g$^{-}-$modes but also sufficiently high
order g$^{+}-$modes become unstable. We provide specific examples
to demonstrate this novel phenomenon that may give rise to several
possibilities to inner stellar core collapses prior to the
emergence of rebound shocks and SN explosions. The time-dependent
factor no longer takes the exponential form but is replaced by a
power-law form.

\subsection{Comparisons with Earlier Model Results}

Conceptually, our perturbation analysis during the phase of
stellar core collapse shows intimate connections to perturbations
before the commencement of stellar core collapse and after the
onset of core bounce.

Physically, the origin of perturbations can be fairly natural for
massive progenitor stars. In terms of stellar evolution and
dynamics, stellar oscillations of p$-$modes, f$-$modes and
g$-$modes may well occur in massive stars, such as red or blue
giants prior to an inner core collapse (e.g. Murphy et al. 2004).
Nuclear burning in massive stars can also provide seed
perturbations (e.g. Bazan \& Arnett 1998; Meakin \& Arnett 2006,
2007a, b). Such pre-existing stellar oscillations and
perturbations associated with nuclear burning serve as sources of
fluctuations during the core collapse phase before the emergence
of a rebound shock around the centre.

GW studied the acoustic stability of core collapse phase for a
conventional polytropic gas and noted that g$-$modes are all
convectively neutral and p$-$modes are all stable. Lai (2000)
extended this polytropic acoustic stability analysis to dynamic
solutions of Yahil (1983) and concluded that unstable acoustic bar
modes (i.e. $l=2$) exist for $\gamma\le 1.09$. The instability is
caused by an insufficient pressure against perturbations due to a
soft EoS and may indicate star formation. A cloud with this
instability tends to deform into an ellipsoid, in which
fragmentation might occur.
Except for this, no other unstable modes relevant to Yahil
solutions were found in Lai's exploration. Lai also claims that
Shu (1977) isothermal EWCS is unstable. In the context of Type II
SNe, he stated that no destructive oscillation modes exist in the
collapsing core before forming a proto-neutron star.
These conclusions are based on the assumption of a conventional
polytropic gas. The fact is that several nuclear processes in SN
explosions most likely make specific entropy distribution
variable.
Then g$-$modes are no longer convectively neutral. In particular,
their stability now becomes sensitive to the evolution of specific
entropy. These unstable g$-$modes should play significant roles
for compact remnants and SNe.

Lai (2000) analyzed and claimed the stability of Yahil (1983)
solution when $\gamma\simeq 4/3$; the background dynamic flows
differ from ours. Lai \& Goldreich (2000) found acoustic
instability growing during dynamic flows. However, their proposed
instability only grows in the supersonic region where the outer
envelope resides (Yahil 1983) while the core still collapses
subsonically.
A main assumption for these results is the same conventional
polytropic EoS for background flow and perturbations. Thus
g$-$modes are convectively neutral in their analysis and only
acoustic p$-$modes exist.


\subsection{Several Aspects of SN Explosions}

One can assess consequences of g$-$mode instabilities during the
phase of the core collapse prior to SNe. Such g$-$mode
instabilities can grow and evolve nonlinearly and several
plausible scenarios may be speculated.

In the presence of a series of core nuclear processes, including
neutronization and electron capture, fluctuations in the specific
entropy during the core-collapse phase is inevitable; in
particular, such unstable g$-$modes, i.e. convective instabilities
destroy the spherical symmetry of a collapsing core before the
emergence of a rebound shock. This is a likely mechanism of
producing asphericity now seemingly necessary for SN explosions.
Two major consequences of such g$-$mode instabilities are as
follows. (i) The formation and proper motion of a proto-neutron
star can be affected. As the collapse becomes aspherical, the
proto-neutron star may also become aspherical. Under favorable
conditions, we speculate that such convective instabilities might
be violent enough to tear a proto-neutron star into pieces.
(ii) Such instabilities are expected to distort the shape of a
rebound shock front. Compared with current numerical simulations,
in which a rebound shock front is initially spherical, then
becomes non-spherical and bends towards a particular direction
where gas is driven out, our proposed non-spherical shock front at
the beginning might give rise to considerable differences for
these numerical simulations.
%
%

There are several conjectures if our proposed instabilities play a
major role in SN explosions. The $l=1$ g$-$mode instability may
give rise to kicks of proto-neutron stars. The $l=2$ instability
may split the central massive core apart and implies a possible
formation of binary pulsars; by adjusting the parameters in our
model, we can have low-order $l=2$ g$-$mode instability in a
central core whose enclosed mass is $\sim 2M_\odot$. High-order
and high-degree g$-$mode instabilities always have a tendency to
tear a central core into pieces. As a consequence, even though
core collapse occurs during which neutronization is triggered and
energetic neutrinos burst out, the outgoing rebound shock may
leave behind a collection of broken clumps. We suggest this
possibility because in observations no central compact object is
found in some SN remnants, including SN1987A (Chevalier 1992;
Manchester 2007; McCray 2009 private communications).

One important prediction of our proposed unstable g$-$mode
convective instabilities before and during the rebound shock
emergence is the resulting turbulent mixing of heavier elements in
the inner layers of the core for SNe. According to the evolution
theory of massive stars (e.g. Nomoto \& Hashimoto 1988; Woosley \&
Weaver 1995), different heavy elements (e.g. Fe, Si, O, C etc.)
lie in ordered core layers from inside out. Without convective
instabilities occurring before the emergence of a rebound shock,
these elemental layers should be more or less kept during a SN
explosion. In this situation, boundaries between elemental shells
are expected to be identifiable. In contrast, for our proposed
g$-$mode convective instabilities, the resulting convective
turbulence with sufficiently fast growths mixes elements Fe, Si,
O, C in different core layers. This convective mixing mechanism of
the core turbulence should bear observational consequences in
detecting various heavy nuclear species and mapping their spatial
distributions in SN remnants.

\subsubsection{Speculations on kicks of radio pulsars}

Observational evidence, including high neutron star peculiar
velocities (e.g. Lyne \& Lorimer 1994; Cordes et al. 1993; Burrows
2000; Arzoumanian et al. 2002;
Hobbs et al. 2005),
the detection of geodetic precession in the binary pulsar PSR
1913+16 (e.g. Wex et al. 2000),
the spin-orbit misalignment (e.g. Kaspi et al. 1996), implies
``kick" processes by which proto-neutron stars gain considerable
kinetic energies during SN explosions. Three major mechanisms have
been pursued (e.g. Lai, Chernoff \& Cordes 2001).
Here, our proposed $l=1$ g$-$mode instabilities in a collapsing
core belong to hydrodynamically driven ``kicks". We shall not
dwell upon the other two, viz. electromagnetically driven ``kicks"
and neutrino-magnetic field driven ``kicks".

The dipole modes of $l=1$, including p$-$modes and both types of
g$-$modes, are characterized by possible displacements of the
central core mass as already noted in stellar oscillations (e.g.
Christensen-Dalsgaard 1976). As such dipole modes involve the core
mass motion about the equilibrium centre, the core mass gains a
certain amount of kinetic energy.
During the violent rebound shock breakout of a SN, this core mass
movement enables the remnant compact object to further gain
kinetic energy and move away from the equilibrium centre. The
overall centre of mass should remain fixed in space.
This is a physically plausible `kick' process. Observationally,
proper velocities of nascent neutron stars are typically $\sim
450\pm 90\ {\rm km\ s^{-1}}$ (e.g. Lyne \& Lorimer 1994) and can
reach as high as $\sim 1600\ {\rm km\ s^{-1}}$ (e.g. Cordes et al.
1993; Burrows 2000).
For all perturbations with $l\neq 1$, the boundary condition at
$x\rightarrow 0$ requires a zero velocity there. In these cases,
the core mass and the equilibrium centre coincide.
Fig. \ref{fig12} offers an example of unstable $l=1$ g$^{+}-$mode.
Unstable $l=1$ g$-$modes of low orders during the core collapse
phase in a massive progenitor star may give rise to the initial
`kick' of a nascent proto-neutron star as joint results of
nonlinear collapse evolution, rebound shock and SN explosion.
Here, the requirement of low-order g$-$modes is to enclose a
sufficient amount of core mass to be kicked out.
Our model results indicate that with a plausible distribution of
specific entropy, low-order $l=1$ g$-$modes can indeed be unstable
during core collapse. Low-order g$-$modes ensures a sufficient
amount of mass inside the inner most node. With such
instabilities, a sizable mass in the central core can be kicked
away from the equilibrium centre.

We may apply this ``kick" scenario to a plausible situation in
which several physical variables are assigned typical values for a
stellar core. From Hillebrandt et al. (1984) and Woosley et al.
(2002), the central coefficient $\kappa_c$ is estimated as $\sim
5\times 10^{14}$ cgs unit. The time $t$ is characterized by the
free-fall timescale $\sim(G\rho)^{-1/2}$ in the order of a few to
several tens of millisecond when the mass density $\rho$ falls in
the range of $\sim 10^{10}-10^{14}{\rm\ g\ cm}^{-3}$. In our
model, the time $t$ is negative (i.e. time reversal) for a
homologously core collapse solution; we thus take the initial time
to be $-50$ ms. Here, $g(x)=1+0.001x^2$ represents a slight
increase of specific entropy with increasing $x$. The case of
$\lambda=0.006$ is close to the maximum value $\lambda_M$ of a
physically allowed $\lambda$ and also leads to a central mass
density of $\sim 10^{10}{\rm\ g\ cm}^{-3}$ consistent with
simulation results (e.g. Bruenn 1985).

\begin{figure}
\includegraphics[width=0.5\textwidth]{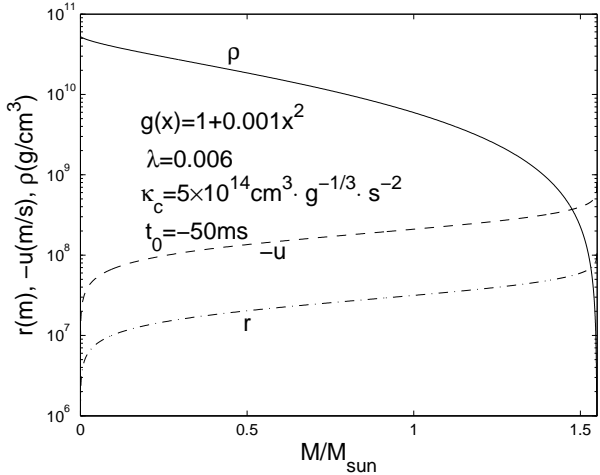}
\caption{
This example presents the initial dynamic background variables for
a homologous core collapse such as the radius $r$, the collapsing
velocity $-u$ and the mass density $\rho$ versus the enclosed mass
$M$ in the unit of the solar mass $M_{\odot}$. }\label{fig13}
\end{figure}

\begin{figure}
\includegraphics[width=0.5\textwidth]{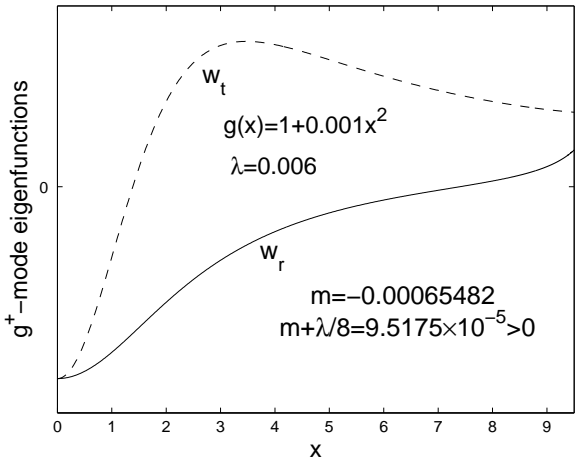}
\caption{
We show an example of the $l=1$ lowest order g$^{+}-$mode.
Parameters for this oscillatory mode are displayed in the figure.
This is a convectively unstable g$^{+}-$mode with $m+\lambda/8>0$.
We can see that the centre of mass has a velocity amplitude, which
contributes to the kick velocity of the core. }\label{fig12}
\end{figure}

Fig. \ref{fig13} illustrates the initial dynamic core collapse of
such a situation, including the radius $r$, the collapsing
velocity $-u$ and the mass density $\rho$ versus the enclosed mass
$M$. Fig. \ref{fig12} shows a $l=1$ dipole unstable g$^{+}-$mode
in which the central core has maximum perturbation amplitudes. It
is estimated that the sound speed of the central core is $(\gamma
P_c/\rho_c)^{1/2}\gsim 10^9 {\rm\ cm\ s}^{-1}$ where the subscript
$c$ indicates physical variables at the centre $x=0$. For a
velocity perturbation only of a few percent of the sound speed, it
can still be several hundred kilometers per second which is
grossly consistent with observed kick velocities of radio pulsars
(e.g. Lyne \& Lorimer 1994). A somewhat larger velocity
perturbation amplitude (say 10\% of the core sound speed) may also
give rise to a kick velocity of $\sim 1600\hbox{ km s}^{-1}$ or
even higher. We propose that this lower-order $l=1$ g$-$mode
instability is a plausible mechanism leading to kicks of central
remnant compact objects in the aftermath of SN explosions.

\subsubsection{Speculations on forming binary pulsars}

A few binary pulsars have been detected, including the famous
PSR1913+16 (Hulse \& Taylor 1975). Ideas have been proposed to
explain their formation (e.g. Flannery \& van den Heuvel 1975).
Our $l=2$ g$-$mode instabilities suggest an alternative yet
plausible origin of such binary pulsars. Unstable $l=2$ g$-$modes
of low orders developed during the stellar core collapse of a
massive progenitor might break the dense core apart before and
during the emergence of a rebound shock and eventually give rise
to binary compact objects. The violent SN explosion drives the two
compact blobs apart. The spins of the two blobs and the binary
orbital motion pick up part of the angular momentum of the massive
progenitor star.
The mass of each components in PSR1913+16 is estimated to be $\sim
1M_\odot$ (e.g. Flannery \& van den Heuvel 1975).
For a central $\kappa_c$ of $\sim 7\times10^{14}$ cgs unit, the
enclosed core mass can be $\gsim 2M_\odot$. That would be a
sufficiently massive core for a possible split into two compact
objects due to nonlinear evolution of unstable $l=2$ g$-$modes and
subsequent core rebounce and SN explosion.

\subsubsection{Speculations on breaking up neutron-rich cores}

We also speculate that under possible and favorable conditions,
the growth of unstable high-order high-degree (e.g. $l\ge 3$)
g$-$modes and their nonlinear evolution might lead to an eventual
breakup of a central proto-neutron star after the rebound shock
emergence and subsequent SN explosion. In this scenario, the
neutronization does occur during the brief core-collapse phase and
high-energy neutrinos escape after a short moment of trapping, but
without forming a coherent central neutron star in the end. The
expected compact remnant is actually shredded into pieces by the
nonlinear evolution of high-order and high-degree unstable
convective g$-$modes. SN1987A might be such an example, for which
no signals from a central compact object have ever been detected
(e.g. Chevalier 1992; Manchester 2007; McCray 2009 private
communications). Unstable high-order high-degree g$-$modes under
other conditions might also lead to the formation of planets
orbiting around a neutron star. These planets can hardly be those
before the SN explosion because they can seldom survive under the
expansion of the progenitor giant and the strong stellar wind as
well as the SN explosion. These high-order high-degree unstable
g$-$modes during the core collapse might help certain amount of
gas concentrated in isolated blobs, leading to the possible
formation of planets around a new-born neutron star. A few neutron
stars with planets have been detected observationally (e.g. Bailes
et al. 1991; Wolszczan \& Frail 1992).

We have revealed and emphasized unstable growths of g$-$mode
convective instabilities during the inner core collapse under
self-gravity inside a massive progenitor star and speculated
several physical consequences of such instabilities about central
compact remnants of SNe. Together with other physical processes at
different stages, such g$-$mode convective instabilities prior to
the emergence of a rebound shock can be a necessary ingredient to
achieve successful SN explosions. Moreover, different modes of
oscillations in our classification provide sources of
perturbations during the emergence of a rebound shock, for
example, the $l=1$ g$-$modes discussed by Burrows et al. (2006,
2007a, b) may have well originated from our $l=1$ g$^{+}-$ and
g$^{-}-$modes during the stellar core collapse phase before the
core bounce. Nonlinear evolution of diverse initial fluctuations
in various stellar core collapse conditions is expected to give
rise to a diversity of possible outcomes for remnant cores of SNe.


\section{Summary and Conclusions}

We have systematically investigated physical properties of 3D
perturbations in a hydrodynamic background of self-similar core
collapse with a general polytropic EoS for a relativistically hot
gas of $\gamma=4/3$, as studied in Lou \& Cao (2008). For both
background and perturbations, the two values of $\gamma$ are taken
to be the same; the case of two different values of $\gamma$ will
be explored separately. Analogous to the mode classification of
stellar oscillations in a non-rotating static star (e.g. Cowling
1941), our 3D general polytropic perturbations are divided into
four distinct classes of modes, viz. p$-$modes, f$-$modes,
g$^{+}-$modes (i.e. with $m<0$) and g$^{-}-$modes (i.e. with
$m>0$), according to their eigenvalue regimes of parameter $m$ in
equation (\ref{pf1}).

Stability properties of these different perturbation modes are
analyzed. Similar to stellar oscillations, acoustic p$-$modes and
f$-$modes remain stable for 3D general polytropic perturbations in
homologous stellar core collapse. This more general conclusion
also confirms the acoustic p$-$mode stabilities claimed by GW
although their p$-$mode eigenvalues appear in systematic errors.
The temporal amplification factor $t^{-1/6}$ in the perturbation
magnitude is associated with the background gas compression during
a homologous inner core collapse. In contrast, g$^{-}-$modes and
sufficiently high-order g$^{+}-$modes are both convectively
unstable modes because the onset criterion of convective
instabilities is now shifted from $m>0$ for a static general
polytropic sphere to $m>-\lambda/8$ for a collapsing general
polytropic core where $\lambda>0$ characterizes the hydrodynamic
background of a homologously collapsing stellar core. The
existence of both types of g$-$modes depends on ${\cal N}^2$, the
square of the Brunt$-$V$\ddot{\rm a}$is$\ddot{\rm a}$l$\ddot{\rm
a}$ buoyancy frequency (see definition \ref{BVfreq}) which is
determined by the evolution of the specific entropy distribution
$g(x)$. Meanwhile, above what value of perturbation degree $l$ the
g$-$modes are unstable also depends on the specific form of
$g(x)$. As an example, the lowest-order $l=1$ unstable
g$^{+}-$mode is shown in Fig. \ref{fig12}. The peak amplitudes of
g$^{-}-$modes lie in regions of ${\cal N}^2<0$, which is analogous
to those in stellar oscillations. These unstable g$-$modes lead to
growths of convective motions in a self-similar collapsing stellar
core. Their divergent growths scale as power laws in time $t$
(with $t<0$) while stable perturbation modes oscillate in the
manner of $\exp(i\zeta\ln t)$.



In analyzing this perturbation problem, we also realize that the
global energy criterion of Chandrasekhar (1939) is not sufficient
to ensure the stability of general polytropic equilibria in view
of the possible occurrence of convective instabilities for
variable entropy distributions (Appendix C).

Compared to possible sources of perturbations proposed in earlier
models, including the so-called ``$\epsilon-$mechanism" before the
onset of inner core collapse and SASI after the core bounce, our
g$-$mode convective instabilities occur during the dynamic core
collapse.
Contrary to earlier theoretical notions, the pre-SN stellar core
collapse phase is most likely convectively unstable due to both
types of g$-$mode instabilities. This is because an exactly
constant specific entropy everywhere in a stellar core should be
extremely rare in any realistic progenitor star (e.g. Bruenn
1985). Therefore, the spherical symmetry of a self-similar
collapsing stellar core should be actually broken up earlier than
presumed by most previous models of SNe.

In our scenario, oscillations of the progenitor star serve as the
most natural source for 3D perturbations. For some regions of
$g'(x)<0$ leading to ${\cal N}^2<0$ locally, the unstable
g$^{-}-$modes, of which low-degree modes may have sufficiently
fast growth rates, will soon dominate and destroy the spherical
symmetry. If $g'(x)>0$ everywhere, high-order unstable
g$^{+}-$modes will grow in a self-similar collapsing core and
evolve nonlinearly.

In the presence of inevitable core g$-$mode convective
instabilities, several possible consequences may follow. Most
prominently, the early break-up of spherical symmetry may lead to
energy concentration in particular directions and there is thus no
need for a rebound shock to push against the entire outer
envelope.
The low-order unstable $l=1$ g$-$modes may give rise to the
initial kick of a remnant central compact object. Meanwhile
unstable g$-$modes correspond to the growth of convective motions
which may stir up heavier elements Fe, Si, O and C in different
inner layers of the collapsing core inside a massive progenitor
via convective turbulence so that a mixed distribution of these
elements to various extents in SN remnants is expected. This
prediction of our model can be tested by nuclear abundance
observations in SNe.

We further suspect that the nonlinear evolution of low-order
unstable $l=2$ g$-$mode instabilities might disintegrate the
central proto-neutron star core into two blobs to form binary
pulsar systems under favorable conditions. Other high-order
high-degree unstable g$-$modes might even prevent the formation of
a coherent central compact object by breaking the core into
multiple pieces. It is speculated that this might happen to
supernova SN1987A without signals from a central remnant compact
object. High-degree and high-order g$-$mode instabilities may lead
to smaller blobs of low masses which might eventually form planets
around a neutron star after a SN explosion.

\section*{Acknowledgements}

We thank the anonymous referee for constructive suggestions to
improve the quality of the manuscript. This research was supported
in part by Tsinghua Centre for Astrophysics (THCA), by the National
Natural Science Foundation of China (NSFC) grants 10373009 and
10533020
at Tsinghua University, and by the SRFDP 20050003088 and
200800030071, the Yangtze Endowment and the National Undergraduate
Innovation Training Project from the Ministry of Education at
Tsinghua University.

\begin{appendix}
\section{Orthogonality of Eigensolutions with Different $m$}

We can now prove that eigenfunctions of different eigenvalues $m$
are mutually orthogonal. In the following illustration, superscripts
$^{(1)}$ and $^{(2)}$ are utilized to distinguish different
eigenfunctions and eigenvalues. Let us proceed to evaluate the
integral of
\begin{eqnarray}
m^{(1)}\int f^3{\bf w}^{(1)}\cdot{\bf
w}^{(2)}dV=-\frac{3}{4}\int\nabla
\left(gf^4\beta_1^{(1)}\right)\cdot{\bf w}^{(2)}dV\nonumber\\
+\frac{3}{4}\int f_1^{(1)}\nabla\left(gf^4\right)\cdot{\bf
w}^{(2)}dV-\int f^3\nabla\psi^{(1)}_1\cdot{\bf w}^{(2)}dV\
.\label{app1}
\end{eqnarray}
The last term on the RHS of eq. (\ref{app1}) leads to
\begin{eqnarray}
\!\!\!\!\!\!\!\!\!\!\!\!\!&&\!\!\!\!\!\!\!\!\!\!\!\!\! -\int
f^3\nabla\psi^{(1)}_1\cdot{\bf w}^{(2)}dV=\int\psi_1^{(1)}
\nabla\cdot\left(f^3{\bf w}^{(2)}\right)dV\nonumber\\
&=&\!\!\!\! -\frac{1}{3}\int\psi_1^{(1)}\nabla^2\psi_1^{(2)}dV
=\frac{1}{3}\int\nabla\psi_1^{(1)}\cdot\nabla\psi_1^{(2)}dV\ .
\end{eqnarray}
The first integration
on the RHS of eq (\ref{app1}) can be re-expressed as
\begin{eqnarray}
&&\frac{3}{4}\int f^4\left(\frac{4}{3}gf_1^{(1)}-\nabla
g\cdot{\bf w}^{(1)}\right)\nabla\cdot{\bf w}^{(2)}dV\nonumber\\
&=&-\frac{3}{4}\int f^3\left(\frac{4}{3}gf_1^{(1)}-\nabla
g\cdot{\bf w}^{(1)}\right)\nonumber\\
&&\qquad\qquad\times\left(ff_1^{(2)}+3\nabla
f\cdot{\bf w}^{(2)}\right)dV\nonumber\\
&=&-\int gf^4f_1^{(1)}f_1^{(2)}dV-3\int f^3gf_1^{(1)}\nabla
f\cdot{\bf w}^{(2)}dV\nonumber\\
&&+\frac{9}{4}\int f^3\left(\nabla f\cdot{\bf
w}^{(2)}\right)\left(\nabla
g\cdot{\bf w}^{(1)}\right)dV\nonumber\\
&&+\frac{3}{4}\int f^4\left(\nabla g\cdot{\bf
w}^{(1)}\right)f_1^{(2)} dV\ ,
\end{eqnarray}
and the second integration on the RHS of eq (\ref{app1}) is equal to
\begin{eqnarray}
3\int f^3gf_1^{(1)}\nabla f\cdot{\bf w}^{(2)}dV+\frac{3}{4}\int
f^4f_1^{(1)}\nabla g\cdot{\bf w}^{(2)}dV\ .
\end{eqnarray}
Note that both $f$ and $g$ depend only on $x$. Therefore, the result
indicates that the integral
\begin{eqnarray}
\int f^3{\bf w}^{(1)}\cdot{\bf w}^{(2)}dV\nonumber
\end{eqnarray}
is manifestly symmetric in terms of superscripts $^{(1)}$ and
$^{(2)}$. Consequently, the orthogonality of eigenfunctions is
proved.

\section{Variational Principle}

The variational principle (e.g. Chandrasekhar 1964) can also be
applied to this eigenvalue problem. Referring to the formula in the
proof of orthogonality in Appendix A, the eigenvalue can be written
as the ratio of two integrals, viz.
\begin{eqnarray}
m=I_2/I_1\ ,\label{defm}
\end{eqnarray}
where
\begin{eqnarray}
I_1&=&\int f^3{\bf w}^2dV\ ,\\
I_2&=&-\int gf^4f_1^2dV+\frac{1}{3}\int|\nabla\psi_1|^2dV\nonumber\\
&&+\frac{9}{4}\int f^3\left(\nabla f\cdot{\bf w}\right)\left(\nabla
g\cdot{\bf w}\right)dV\nonumber\\
&&+\frac{3}{2}\int f^4f_1\left(\nabla g\cdot{\bf w}\right)dV\ .
\end{eqnarray}
Using the variational principle, we will show that the eigenvalue
parameter $m$ has a stationary property when $I_1$ and $I_2$ are
evaluated in terms of the true proper solutions. According to
equation (\ref{defm}),
\begin{eqnarray}
\delta m=
\left(\delta I_2-m\delta I_1\right)/I_1\ ,
\end{eqnarray}
where $\delta I_1$ and $\delta I_2$ are the changes in $I_1$ and
$I_2$ in response to the variation $\delta {\bf w}$ in ${\bf w}$. We
have
\begin{eqnarray}
\delta I_1=2\int f^3{\bf w}\cdot\delta{\bf w}dV\ ,\label{deltaI1}
\end{eqnarray}
and
\begin{eqnarray}
\delta I_2\!\!\!\!\!\!&=&\!\!\!\!\!\!-2\int gf^4f_1\delta f_1
dV+\frac{2}{3}\int\nabla\psi_1\cdot\nabla\delta\psi_1dV\nonumber\\
\!\!\!\!\!\!&&\!\!\!\!\!\!+\frac{3}{2}\int f^4\left(\nabla
g\cdot{\bf w}\right)\delta f_1dV+\frac{3}{2}\int f^4f_1\left(\delta
g\cdot\delta{\bf w}\right)dV\nonumber\\
\!\!\!\!\!\!&&\!\!\!\!\!\!+\frac{9}{4}\int f^3\left(\nabla
f\cdot\delta{\bf w}\right)\left(\nabla g\cdot{\bf w}\right)dV\ .
\end{eqnarray}
Keeping in mind of equations (\ref{pf1})$-$(\ref{pf4}), we can write
the variation $\delta I_2$ as
\begin{eqnarray}
\delta I_2=2\int\left\{-\frac{3}{4}\left[\nabla\left(gf^4
\beta_1\right) \right.\right.\qquad\qquad\qquad \nonumber\\
\left.\left. -f_1\nabla\left(gf^4\right)\right]-f^3\nabla\psi_1
\right\} \cdot\delta{\bf w}dV\ .\label{deltaI2}
\end{eqnarray}
By equations (\ref{deltaI1}) and (\ref{deltaI2}), it then follows
that $\delta m=0$ if
\begin{eqnarray}
mf^3{\bf w}=-\frac{3}{4}\left[\nabla\left(gf^4\beta_1
\right)-f_1\nabla\left(gf^4\right)\right]-f^3\nabla\psi_1\ ,
\end{eqnarray}
which is precisely equation (\ref{pf1}) for the eigenvalue problem
of our model formulation.

\section{\quad Total Energy of a\\ \qquad\ General Lane-Emden Sphere}

In the model formulation of Lou \& Cao (2008), the situation of
$\lambda=0$ represents the limiting static general polytropic
sphere. For the conventional polytropic equation of state (i.e. a
constant specific entropy everywhere independent of time $t$), the
governing equation of the static equilibrium returns to the
well-known Lane-Emden equation (e.g. Eddington 1926; Chandrasekhar
1939). For a general polytropic gas sphere with a variable radial
distribution of the specific entropy, we refer to the governing
equation with $\lambda=0$ as the general Lane-Emden equation (see
footnote 1 in the main text). We emphasize that the linear stability
property now depends on the radial distribution of the specific
entropy. If the specific entropy decreases with radius $r$,
corresponding to a decreasing $g(x)$, convectively unstable
g$^{-}-$modes will develop and drive the static sphere out of
equilibrium configuration. The total energy that is equal to the
gravitational energy plus the thermal energy together still remains
zero for $\gamma=4/3$; however, this situation under the
conventional polytropic assumption is referred to as being
marginally stable by Chandrasekhar (1939).

The hydrostatic equilibrium equations are
\begin{eqnarray}
\frac{dM}{dr}=4\pi r^2\rho\ ,\label{Ss1}\\
\frac{dP}{dr}=-\frac{GM\rho}{r^2}\ .\label{Ss2}
\end{eqnarray}
Using equation (\ref{Ss1}) to eliminate
$M$ in equation (\ref{Ss2}) and assuming the radial distribution
of $P/\rho^{\gamma}$, one readily arrives at the dimensional
general Lane-Emden equation whose dimensionless form appears as
equation (\ref{equf}) with $\lambda=0$.

The total energy of the system can be derived as follows. The total
gravitational energy is
\begin{eqnarray}
E_G=-\int \frac{GM\rho}{r} 4\pi r^2dr=\int\frac{dP}{dr} 4\pi r^3dr\
\end{eqnarray}
and the total thermal energy is
\begin{eqnarray}
E_T=\int\frac{P}{(\gamma-1)\rho} 4\pi r^2 \rho
dr=\frac{4\pi}{(\gamma-1)}\int Pr^2dr\ .
\end{eqnarray}
The total energy of a general Lane-Emden sphere is
\begin{eqnarray}
E=E_G+E_T=-(3\gamma-4)E_T\ ,
\end{eqnarray}
where finite pressure at the centre and zero pressure at the
stellar surface are presumed. Consequently, we prove that
irrespective of the radial distribution of the specific entropy,
the total energy of a general polytropic Lane-Emden sphere remains
zero for
$\gamma=4/3$. Evidently, this does not
guarantee the linear stability of such a equilibrium
configuration.

The simple total energy criterion is not enough to
judge the stability of the equilibrium configuration. The statement
of Chandrasekhar (1939) is only valid because the assumption of a
conventional polytropic gas excludes g$-$modes, leaving them as
convectively neutral. In fact, linear stability properties rely on
the radial distribution of the specific entropy. For instance, we
find in this paper g$^{-}-$mode instability for a static general
polytropic Lane-Emden sphere with the square of the
Brunt-V$\ddot{\rm a}$is$\ddot{\rm a}$l$\ddot{\rm a}$ frequency
${\cal N}^2<0$ in some parts.
We therefore conclude that simply using the polytropic index
$\gamma$ to judge the stability property of a static sphere is not
enough for a general polytropic Lane-Emden sphere. One still needs
to perform a detailed linear stability analysis for such a
hydrostatic equilibrium configuration.
\end{appendix}
\end{document}